\documentclass{aa}

\usepackage{graphicx}
\usepackage{txfonts}
\usepackage{booktabs}
\usepackage{longtable}

\usepackage{subfigure}
\usepackage{xcolor}

\usepackage[]{hyperref} 

\bibpunct{(}{)}{;}{a}{}{,} 

\makeatletter
\renewcommand*\aa@pageof{, page \thepage{} of \pageref*{LastPage}}
\makeatother

\begin{document}

   \title{TT Arietis: New approach to the analysis of quasi-periodic oscillations\thanks{Corresponding authors; ignacio.vega@postgrado.uv.cl; nikolaus.vogt@uv.cl; andrea.rojas.astron@gmail.com}}

   \subtitle{}

   \author{I. Vega-Manubens
          \inst{1}
          \and
          N. Vogt\inst{1}
          \and
          A. Lopera-Mejía\inst{1}
          \and
          G. Aravena-Rojas\inst{1,2}
          \and
          P. A. Rojas Lobos\inst{3}
          }

   \institute{Instituto de Física y Astronomía,
Universidad de Valparaíso, Av. Gran Bretana 1111, Valparaíso, 2360102, Chile
        \and
        Vera C. Rubin Observatory, La Serena, Chile
        \and
        Universidad Andres Bello, Facultad de Ciencias Exactas, Departamento de Ciencias Físicas, Instituto de Astrofísica, Av. Fernández Concha 700, Santiago, Chile
             }

   \date{Received ; accepted}

  \abstract
   {TT Arietis (TT Ari) is a nova-like cataclysmic variable of the VY Scl subtype with light-curve variations on multiple timescales. In addition to the superhump modulation, quasi-periodic oscillations (QPOs) have been found.}
   {Our aim is to determine the occurrence, strength, and variability of QPOs in TT Ari based on more complete data than in previous works. }
   {The data were obtained during the high state of TT Ari in October 2012 by the MOST space telescope, covering a total of 361.2 hours of continuous observation. We searched for frequencies over subsets of time using a Fourier-like power spectrum and then added the frequencies together, forming groups.}
   {Our method has revealed QPOs that occur in `frequency groups', which are events with a continuous oscillation of similar, constant or slowly variable frequency. We report a total of 160 frequency groups in the period range between 14 and 53 minutes (27 and 98 days$^{-1}$), with two peaks in the power spectrum at 18.5 and 33.8 minutes (42.5 and 77.5 days$^{-1}$). The duration of these frequency groups varies between 0.72 and 7.5 hours (average 2.8 hours) revealing between 3 and 18 complete cycles in the light curve. Most of them show significant frequency variations over the course of their duration. Sometimes two frequency groups occur simultaneously. An analysis with randomised data confirms that stochastic processes can only explain a fraction of the QPOs found. The occurrence of QPOs appears not to be related to the superhump phase.}
   {}


   \maketitle

\section{Introduction}
\label{sec:intro} 
Cataclysmic variables (CVs) provide the opportunity to study different mass-transfer and accretion conditions within the same binary, which consists of a white dwarf primary, with its surrounding accretion disk, and a Roche-lobe-filling red dwarf as the secondary component.

Compared to other accreting objects, the dynamical timescale on which accretion proceeds in CVs is relatively short and thus more tractable; CVs are natural laboratories for accretion-disk physics. Well-known targets for such studies are dwarf novae, due to their eruptions at typical timescales of days to months, but nova-like CVs of the so-called VY Scl subtype also show dramatic variations, fading occasionally by several magnitudes on timescales of between months and years.  TT Arietis (TT Ari) is an example of this subclass, and is one of the brightest CV stars on the sky. It is most often found in a `high state' at V $\approx$ 10.5$^{m}$ but can be as faint as 16$^{m}$ in its `low state'. Throughout the history of observations of TT Ari, these low states have repeated every 20 to 25 years and can last between 500 and more than 2000 days, including the decline and rise phases. Recently, during a high state, the light curve of TT Ari was found to be dominated by `negative superhumps' \citep{1999dicb.conf...61P} in a nearly sinusoidal form with slightly variable periods that are about 3.3\% shorter than its orbital period of 0.13755 days. For instance, in 2007, a superhump period of 0.1331 days was observed \citep{Vogt_2013}. A review of the general behaviour of TT Ari during the past few decades is given by \cite{Bruch_2019}. 

The present study is dedicated for the most part to investigating the quasi-periodic oscillations (QPOs) superimposed on the superhump light curve of TT Ari. A comparison between previous investigations into QPO periods is presented in Table \ref{tab:table1}. 

When trying to study QPO behaviour in detail, Earth-bound observations often suffer from necessary gaps in the coverage. The Canadian space telescope Microvariability and Oscillations of Stars (MOST) offers the unique possibility to avoid this problem, due to its uninterrupted observing mode over several weeks. Here we describe almost uninterrupted observations over a total of 361 hours, reporting and discussing details of the frequency, duration, and amplitude of significant QPOs with unprecedented completeness.
\begin{table*}
\centering
\caption{QPO frequencies for TT Ari.}
\label{tab:table1}
\begin{tabular}{llllll}
\hline
Observation  & Observation details & Passband/ & $f$ peaks  & $f$ range& References\\
Year &  & filters & (d$^{-1}$) &  (d$^{-1}$)  &\\
\hline
1961--1962 & \begin{tabular}[c]{@{}l@{}}1 observing run (3 hours) \\ of observations.\end{tabular} & UV & \begin{tabular}[c]{@{}l@{}}34.1, 81.8 \\ and 103.6\end{tabular}  &... & 1\\[0.3cm]
\begin{tabular}[c]{@{}l@{}} 1961, 1962\\and 1966\end{tabular}& \begin{tabular}[c]{@{}l@{}}15 observing runs, \\ each of 2--5 hours.\end{tabular} & V and U & ... &36--140& 2  \\[0.3cm] 
1961--1985 & 30 observing runs. & \begin{tabular}[c]{@{}l@{}}Varied (U,B,\\ UB, UVB)\end{tabular} & \begin{tabular}[c]{@{}l@{}}53.3  in 1961, up\\to 84.7  in 1985\end{tabular}  &... &  3 \\[0.3cm]
1987--1988 & 12 nights of observation. & Filter V &  72  &65.4--80  & 4 \\[0.3cm]
1988 & \begin{tabular}[c]{@{}l@{}}3 nights. Around 250 \\ observations per night.\end{tabular} & 3250--5500 Å & 94.1 &... & 5 \\[0.3cm]
1988 & 16 nights. & \begin{tabular}[c]{@{}l@{}}5 bands, but \\ mostly B\end{tabular} &\begin{tabular}[c]{@{}l@{}}27.9, 52.4, 60.7\\ and 101.4\end{tabular}  &...  & 6 \\[0.3cm]
1987--1998 & \begin{tabular}[c]{@{}l@{}}45 observing runs. Over  \\  600 measurements.\end{tabular} & \begin{tabular}[c]{@{}l@{}}Standard UBV\\ system\end{tabular} & ... & 41.1--120 & 7 \\[0.3cm]
1994 & \begin{tabular}[c]{@{}l@{}}16 nights (258 hours)    \\ of observation, \\51\,299 data points.\end{tabular} & Close to B & 48 and 68.5 &24--139 &  8 \\[0.45cm]
2005 & \begin{tabular}[c]{@{}l@{}}51 observational runs,  \\ covering 226 hours.\end{tabular} & Filter R & 66.7 (mean)  &... & 9 \\[0.3cm]
2012 & \begin{tabular}[c]{@{}l@{}}15 days (361 hours) of \\continuous observation \\and 21\,166 data points.\end{tabular} & 3500--7500 Å & 42.5 $\&$ 77.5 &27--98 & 10 \\ 
\hline
\end{tabular}
\tablefoot{
Most of the presented studies involve a periodogram analysis and some sort of pre-whitening of unwanted periods. Details about the methods are found in the articles themselves. Columns 4 and 5 show information about frequencies ($f$) related to QPOs: $f$ peaks refers to specific frequencies found as peaks in the periodogram, while $f$ range presents frequency ranges in which QPOs are more common.
}
\tablebib{
(1) \citet{William_1966}; (2) \citet{2014AcA....64..167S} ; (3) \citet{Semeniuk_1987} ; (4) \citet{1988AcA....38..315U} ; (5) \citet{Hollander_1992}  ; (6) \citet{Tremko_1996} ; (7) \citet{Kraicheva_1999} ; (8) \citet{Andronov_1999} ; (9) \citet{Y.Kim}; (10) Dominant frequencies from this work.
}
\end{table*}

\section{Methodology}
\subsection{The data}
\label{sec:The data} 

The observations were carried out in the V filter, and were obtained from the Canadian space telescope MOST.
The telescope  was orbiting at an altitude of 820 km, with a period of 101 minutes \citep{2011ApJ...735...34C, 2003PASP..115.1023W, 1999JRASC..93..183M}, and was equipped with a Rumak-Maksutov 15 cm telescope. Its filter covers a wavelength range from 3500 Å to 7500 Å. The telescope has a limiting magnitude of 12$^{m}$ in the visual range, and is equipped with a photometric CCD.

\begin{figure}
        \includegraphics[width=\columnwidth]{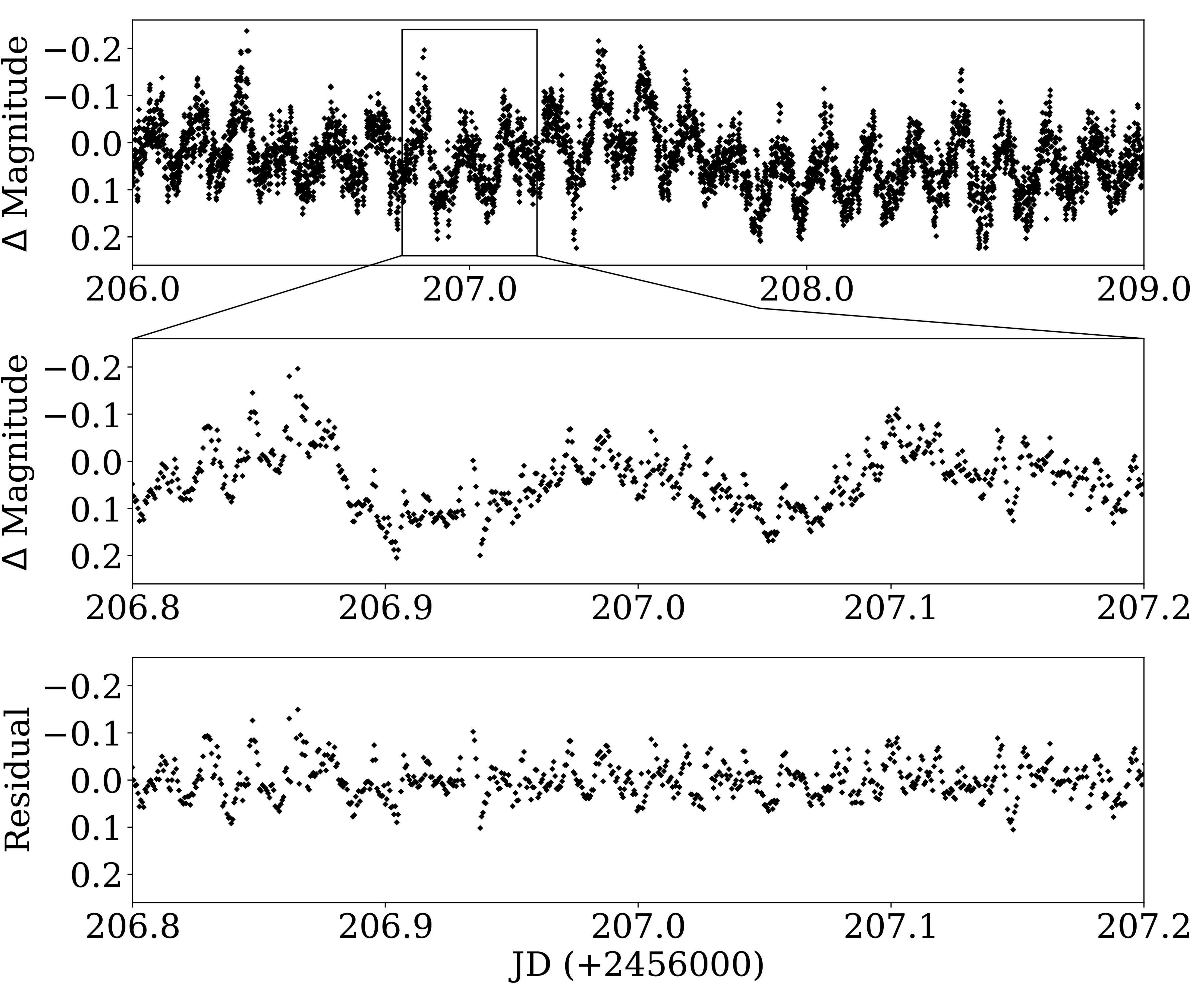}
    \caption{Light-curve segment of TT Ari, showing a typical time interval in its high state. The top panel shows a  three-day span from October 2012, while the middle panel shows segment of  a 0.4 days (9.6 hours) within the same time interval. The bottom panel shows the residual data from the spline interpolation. Removing the superhump modulation reveals other brightness variations, including QPOs.
}
    \label{fig:figure1}
\end{figure}

The data correspond to the high state of TT Ari in October 2012, with a magnitude of between 10.8 and 10.4, obtained from the AAVSO site \citep{aavso}. These data cover 15 days (361.2 hours) with a time resolution of 60.38 seconds and include 21166 individual data points. Observations were almost continuous, with very small gaps in between continuous observation runs; these intermissions in observations are detailed as follows: There are several two-minute gaps, which add up to 3.6\% of the sample. There are also intervals of between three and seven minutes that together do not reach 1\%. Finally, there are five gaps of longer than 45 minutes that together cover about 6\%, leaving a total observation coverage of 89.5\%. The longer gaps are in the regions of JD [$211.3 - 211.8$] and [$214.5 - 215.7$]. Here and hereafter, JD refers to Julian date +2\,456\,000.  Details of the observations can be found at Rojas Lobos et al. (in prep.). The MOST data for TT Ari are in the form of a time-resolved photometric light curve. The data points have a magnitude range of 0.4627, with a mean error of 0.0068. An example light curve is presented in Fig. \ref{fig:figure1}.

\subsection{Power-spectrum analysis and frequency groups}
\label{sec:frequency groups}
The dominant frequency in the light curve of TT Ari, the superhump, is removed with a spline interpolation. Our code also averages out small differences in magnitude over intervals of one day. After the removal of the superhump, the mean magnitude of the residual is set to zero for convenience. The residual has variations with a mean half amplitude of 0.11 in magnitude, with some outliers reaching up to 0.20. This process can be seen in the bottom panel of Fig. \ref{fig:figure1}.

The residuals were divided into segments of 0.1 days (d). Each segment has a 0.09 d overlap with the subsequent segment, resulting in a total of 1489 segments, which were analysed individually with the Lomb-Scargle (LS) periodogram (\cite{Lomb}, \cite{Scargle}). A Fourier-like power spectrum was used to obtain the dominant frequencies ($f$). 

The frequency interval between 20 and 140 d$^{-1}$ was chosen because other authors reported strong peaks in the power spectrum in similar ranges: \cite{Tremko_1996} ran periodograms between 0 and 250 d$^{-1}$ and found most of the peaks in the 28--103 d$^{-1}$ range. 
Searching for frequencies in the 10--2500 d$^{-1}$ interval, \cite{Andronov_1999} found the highest peaks in the power spectra in the 24--139 d$^{-1}$ range. Furthermore, \cite{Kraicheva_1999} analysed light curves using the method of \cite{Scargle}
in the 0--300 d$^{-1}$ interval, and found many equally strong power peaks in the range of 40--120 d$^{-1}$.

After subtracting the superhump effect, the next step was to find remarkable frequencies. For this, we used the Lomb-Scargle (LS) periodogram from the Astropy series \citep{VanderPlas_2012, VanderPlas_2015} with the `psd' normalisation option. This allows direct comparison of power values across different segments and frequencies. The resulting frequencies from the power spectrum went through a two-step filtering and grouping analysis. In the first filter, we removed frequency peaks with low power, leaving only the most prominent frequencies in each segment of  0.1 days. To accomplish this, we set a threshold at 10\% of the normalised frequency power units, which means that any peak below the value of 0.1 was considered as noise. As for the grouping, our goal is to determinate which frequencies survive the longest. The permanence of a single frequency is defined by the number of individual segments whose frequency (obtained by LS) is similar to the frequencies of the adjacent segments. Each nearby segment with a similar frequency is added together to a group, and when the peak frequencies from LS change drastically, the group ends. Hereafter, these are referred to as `frequency groups'. All resulting frequencies that make it through the first filter are either grouped into a frequency group or dropped. 

\begin{figure}
\begin{subfigure}{}
        \includegraphics[width=\columnwidth]{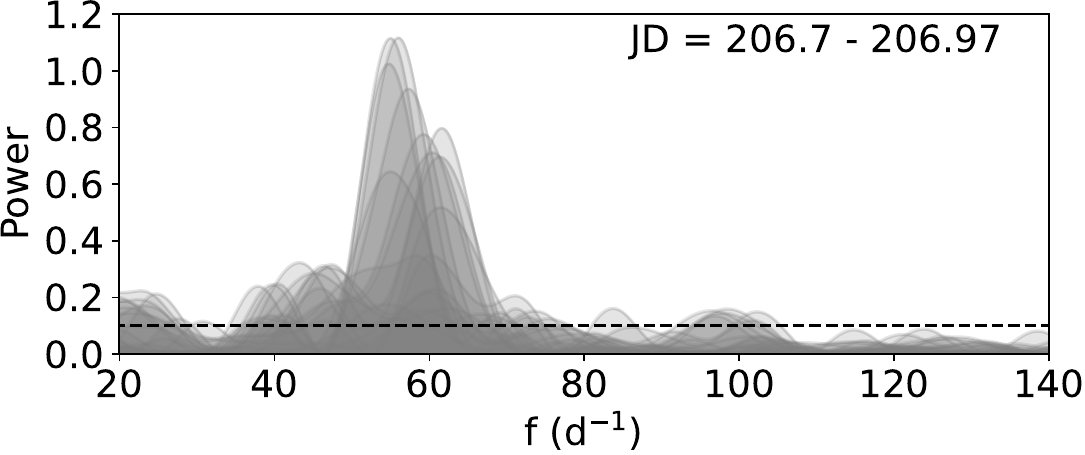}
    \end{subfigure}
    \vspace{-1.2\baselineskip}

\begin{subfigure}{}
        \includegraphics[width=\columnwidth]{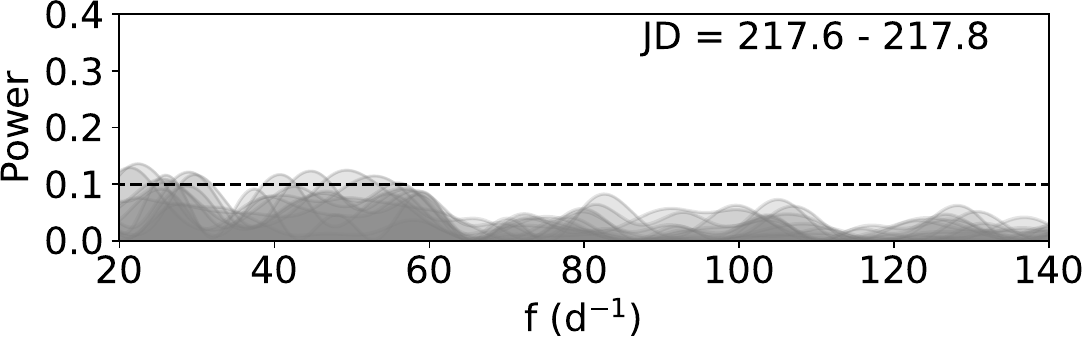}
    \end{subfigure}

\caption{
Power spectrum of the frequency range, calculated for certain time periods. The dashed line corresponds to the power threshold of 0.1. Top panel: Period with one QPO. This corresponds to a specific frequency group visible in Fig. \ref{fig:figure3} at starting date JD 206.7. Bottom panel: Time period without QPOs, appearing as a gap in Fig. \ref{fig:figure3} around JD 217.7.
}
\label{fig:figure2}
\end{figure}

\begin{figure}

\begin{subfigure}{}
\includegraphics[width=1.0\linewidth]{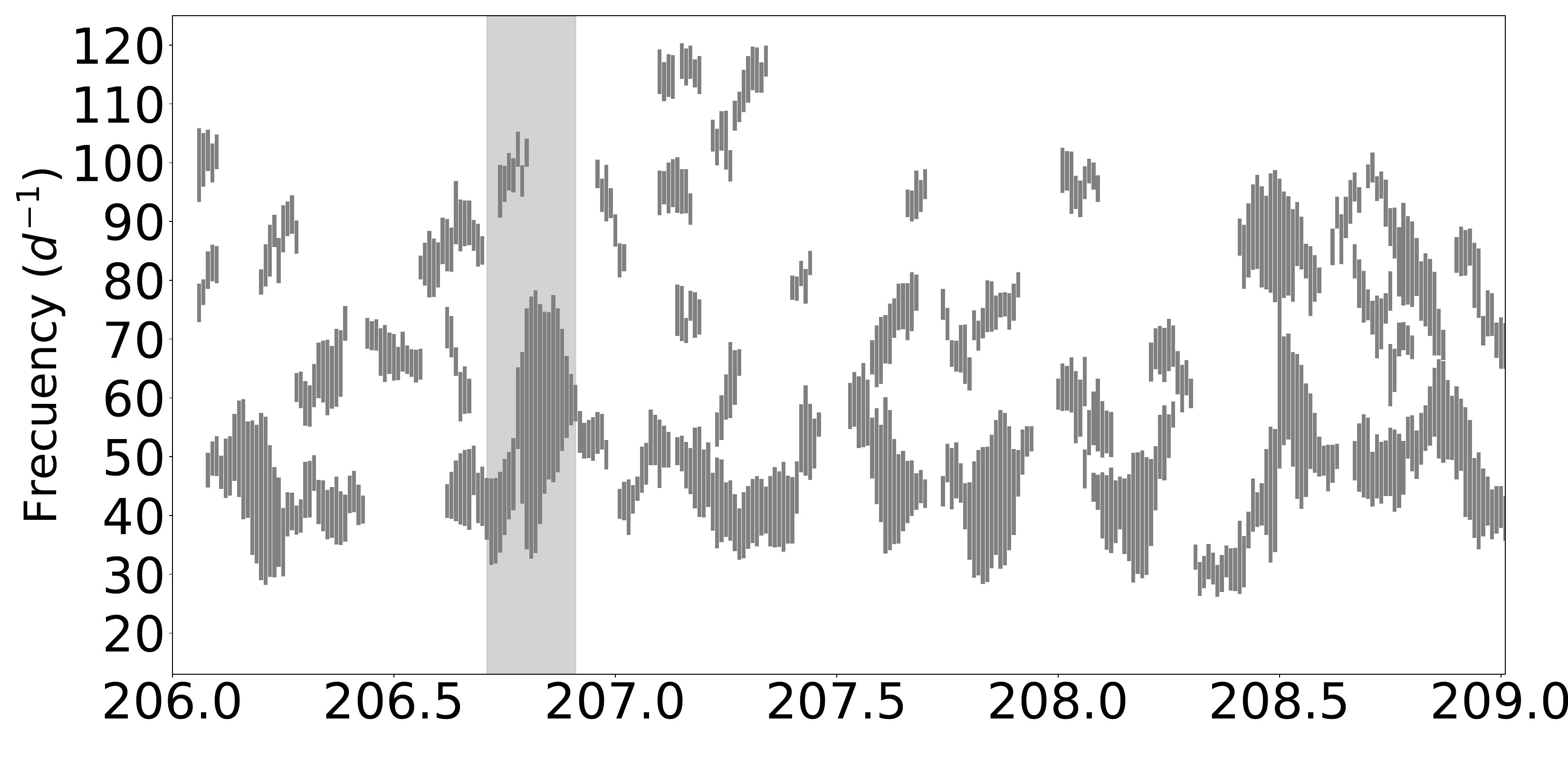}
\end{subfigure}
\vspace{-1.5\baselineskip}

\begin{subfigure}{}
\includegraphics[width=1.0\linewidth]{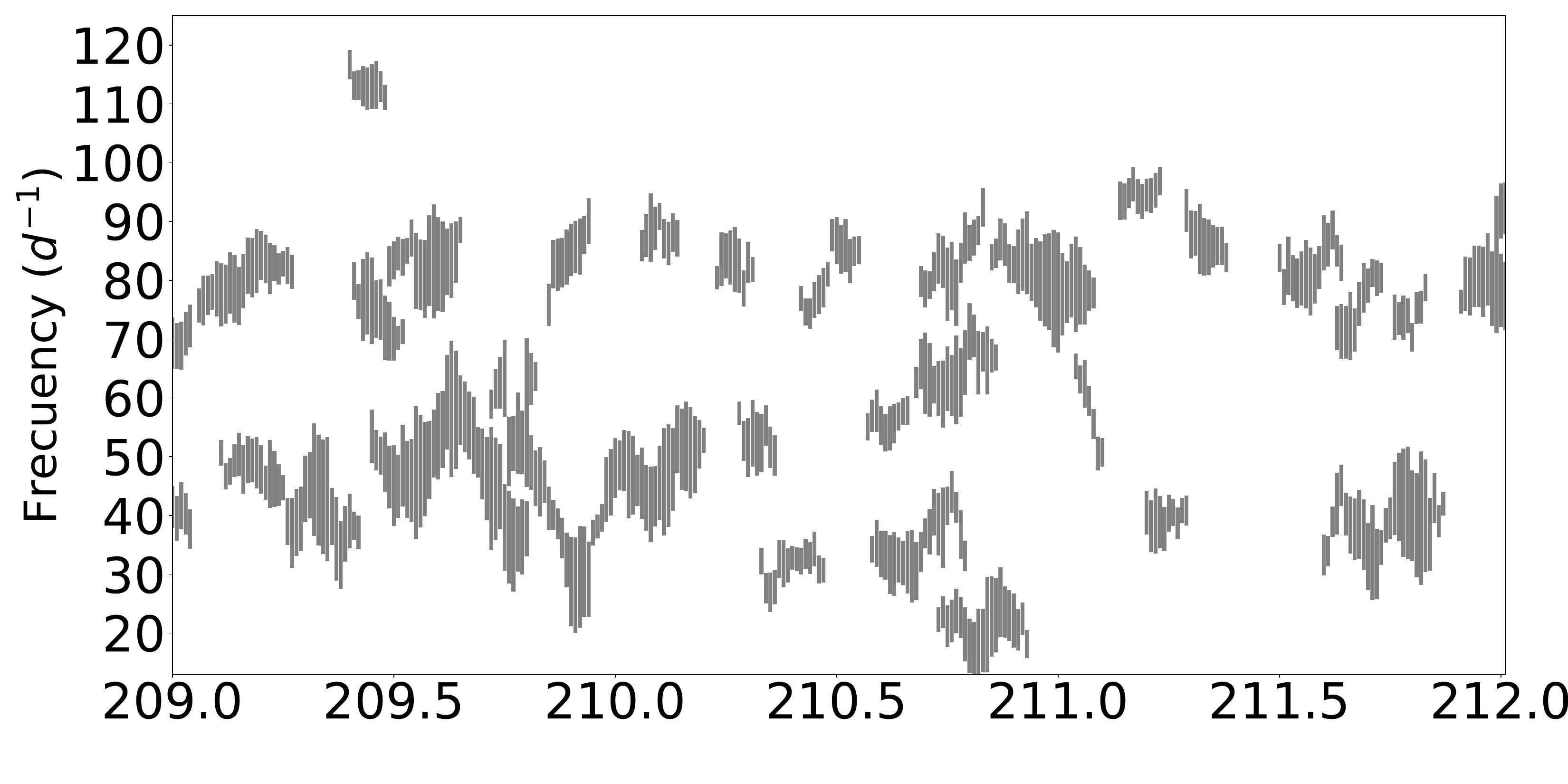}
\end{subfigure}
\vspace{-1.5\baselineskip}

\begin{subfigure}{}
\includegraphics[width=1.0\linewidth]{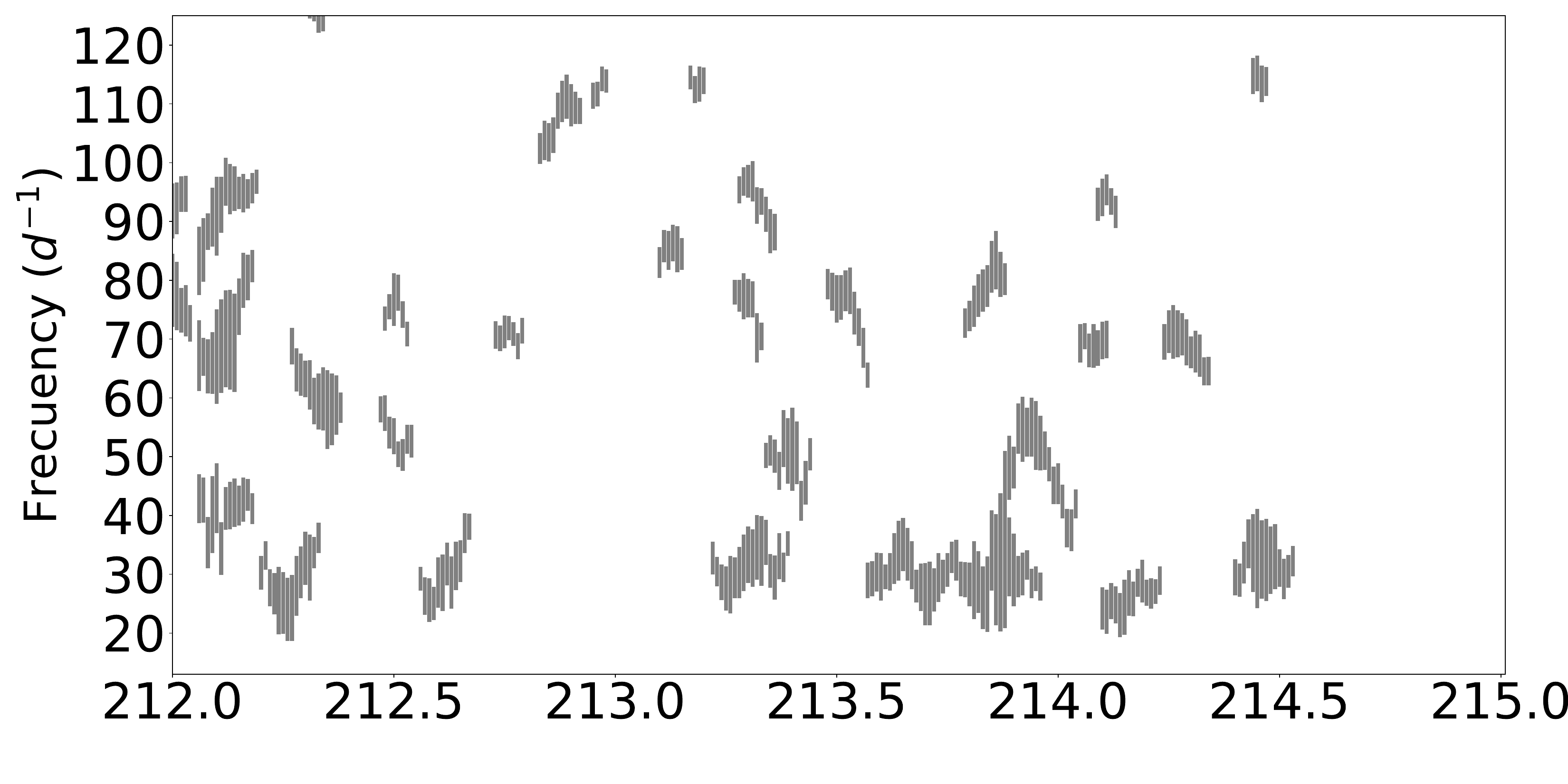}
\end{subfigure}
\vspace{-1.5\baselineskip}

\begin{subfigure}{}
\includegraphics[width=1.0\linewidth]{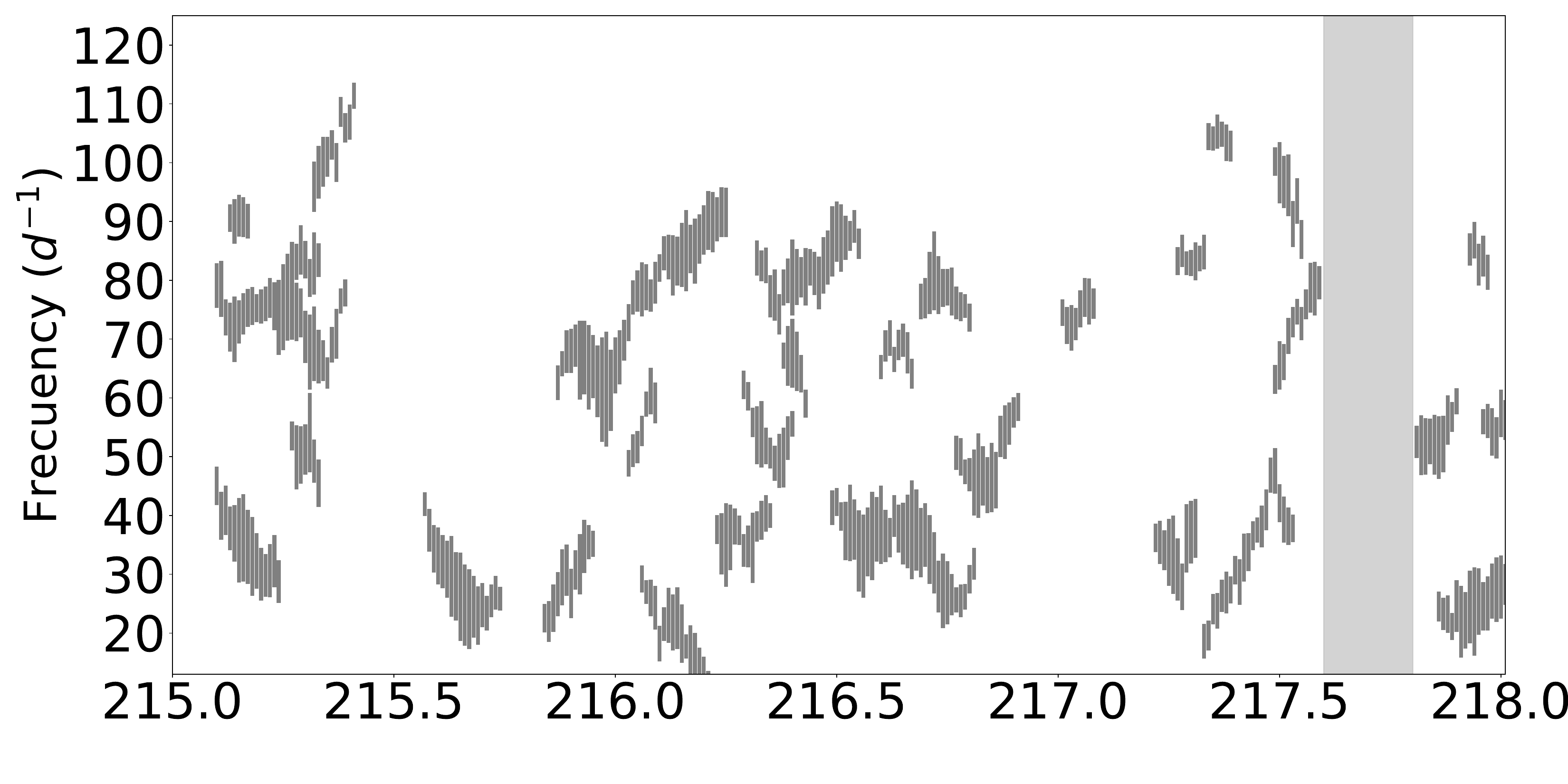}
\end{subfigure}
\vspace{-1.5\baselineskip}

\begin{subfigure}{}
\includegraphics[width=1.0\linewidth]{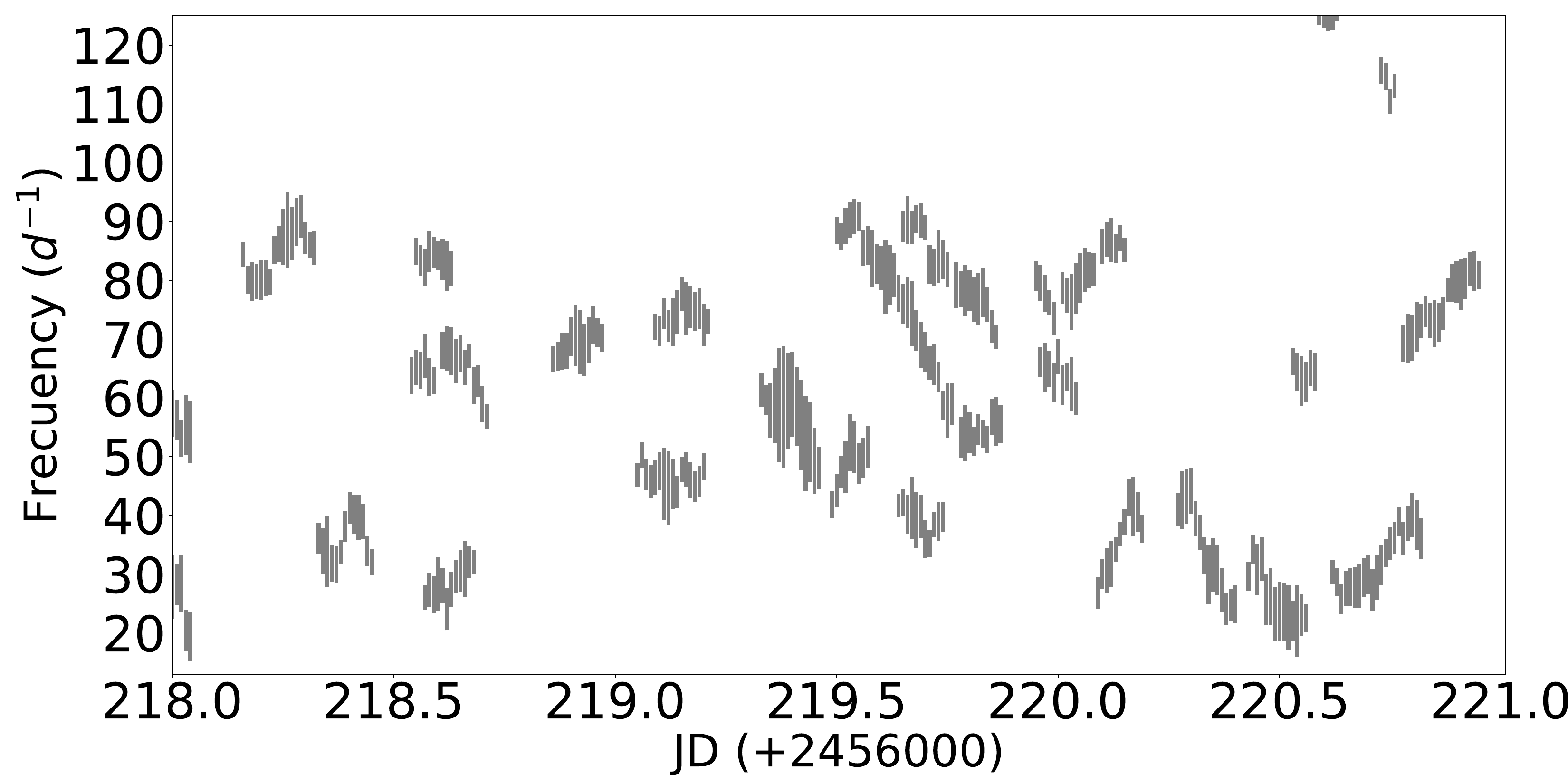}
\end{subfigure}
\caption{Frequency groups throughout the entire 15 days of observations. Each panel corresponds to a three-day segment. Each line in a group represents a frequency peak in the power spectrum, as explained in sect. \ref{sec:frequency groups}. The light grey region at JD 206.7 corresponds to the power spectrum analysed in Fig. \ref{fig:figure2} (top panel), while the filled region at JD 217.7 represents the bottom panel of Fig. 2.
}
\label{fig:figure3}
\end{figure}

Figure \ref{fig:figure2} shows the power spectrum of the frequency range from 20 to 140 d$^{-1}$. Each curve represents the spectrum for a small time period, with a peak at the zone with the highest power. To create frequency groups, we select the five strongest peaks over the power threshold of 0.1 (dashed line). Only frequencies with adjacent peaks are considered; isolated frequencies are excluded. In the top panel, we see that there are numerous peaks above 0.1, and they form a frequency group around 60 d$^{-1}$, which can also be seen in Fig. \ref{fig:figure3} in the filled area starting at JD 206.7. Meanwhile, in the bottom panel, we see a segment with no active QPOs. Here, some peaks reach above 0.1, but they are singular cases with no nearby frequencies. Therefore, they cannot  be grouped and are dropped by the filters. This time period can be seen as a gap in Fig. \ref{fig:figure3} in the filled area around JD 217.7. 

In order to search for stable frequencies, a second filter is applied directly on the groups. In this filter, frequency groups with short duration are removed, as are those with a small number of cycles. These qualities are defined as follows: The duration is defined by the number of adjacent frequencies that form a group. For a group to be considered as valid it must be formed by at least three frequencies; since we compute frequencies in intervals of 0.01 days (0.24 hours), the minimum duration of a group is 0.72 hours. The number of completed cycles ($N_{c}$) is defined as the ratio between the duration (in hours) of the frequency group and the period obtained from the LS algorithm. As an example, a group with a mean frequency of 48 d$^{-1}$ (which is equivalent to a period of 0.5 hours) and a duration of 2 hours can complete at least 4.0 cycles during the observation period. We set the minimum $N_{c}$ as 3.0, as this value gives us a useful number of well-defined frequency groups.

There are other two quantities being tracked: the frequency delta ($f_{\Delta}$) and the frequency amplitude ($f_{A}$). $f_{\Delta}$ measures the difference in frequency between the last and first frequencies in each group, and $f_{A}$ is the difference between the higher and lower frequencies in each group. They can be seen in Tables \ref{tab:table2}, \ref{table3_new}, and \ref{tab:table3}.

We believe the procedures applied to the data (spline interpolation, filtering, and grouping) minimise cases of isolated frequencies with high power but low duration, which are not relevant to the aims of the present study. Our hypothesis is that any frequency corresponding to a real periodicity in the star will have permanence in time.

In Fig. \ref{fig:figure3}, the frequencies are represented as vertical lines whose length is proportional to the power obtained from the LS power spectrum. The factor of proportionality was chosen to prioritise a better visualisation of the data, where the higher the power, the longer the line length. Whenever two vertical lines overlap, they are considered to belong to the same group. 

\subsection{Validation with synthetic light curves}
\label{sec:random_data}

Alongside QPOs, TT Ari exhibits flickering, which is a process of random magnitude fluctuation. Since it is difficult to isolate the flickering (as explained in sect. \ref{sec:flick}), we instead quantify the effects of random fluctuations. To this end, we need to measure the influence of stochastic processes on the frequency groups. One way to quantify this is to create synthetic light curves and record how many frequency groups appear after performing the same frequency analysis as that performed on the original light curve. Groups found in synthetic light curves count as `false positives'; frequencies that do not correspond to QPOs and are simply the result of the superposition of random data points.

These synthetic light curves are created by shuffling segments of the original light curve. Each data point inside each segment gets a new, random residual from the original light curve, while its JD is kept the same. We shuffle segments of the observation instead of individual data points because single observations are not independent of one another; because of flickering, data points are dependent on one another over short time intervals, length of which we need to estimate. Using an autocorrelation function of the original light curve, we obtain a full width at half maximum (FWHM) of four data points. Using this number, we conclude that segments of the light curve containing four data points can be considered as independent of one another. Then, all segments are randomly shuffled while maintaining the inner order of their data points. This generates a new, synthetic light curve, which is then filtered by the same criteria and parameters as described in sect. \ref{sec:frequency groups}. We are interested in the number and size of frequency groups for these randomised light curves.
We excluded the regions JD [$211.3 - 211.8$] and [$214.5 - 215.7$], as the gaps in our observations could produce artificial features.
The shuffling process was repeated 100 times, adding together all frequency groups found. Our findings regarding these randomised data sets are presented in sect. \ref{sec:random_data_findings}.

\section{Results}
\label{sec:results}

\subsection{Frequency group distribution}

After applying our method and using a filter of power $P \ge$ 0.10, we found a total of 1316 frequencies, forming 160 frequency groups. Each of the individual groups and their relevant quantities are shown in Table \ref{tab:table3}. A second, smaller set of 80 frequency groups was obtained using a stricter filter of $P \ge$ 0.15. These groups are presented in Table \ref{tab:table4}.

Figure \ref{fig:figure3} shows the behaviour of frequency groups over time in five panels, each comprising a three-day segment of observations. The first panel contains the first segment, where we find 40 frequency groups, while the following four segments contain around 30 groups each. This drop in group count may be caused by the small intermissions in observation discussed in sect. \ref{sec:The data}, which are more common in the segments in panels two to five. An important detail is that the first two segments have groups with stronger mean power compared with the last three: the average $\overline{P}$ for groups in each of the five segments was, respectively, $0.21, 0.20, 0.17, 0.18$, and $0.16$. Besides these small differences between panels, frequency groups are evenly distributed over the 15 days of observations. We see no common repeating pattern of the frequency groups, but some groups appear at two different frequencies at the same time. These groups are more numerous near certain frequency ranges, namely around 40 and 75 d$^{-1}$. These frequency ranges are presented in more detail in sect. \ref{sec:frequency ranges}. Frequency groups have variable size, duration, and slope, with some being constant in time and others presenting drastic frequency changes. Longer groups are of special interest as they can show some of the following characteristics: peaks and valleys, and separation into two groups or convergence into a single frequency. Frequency groups are present for a significant part of the observation time, and on certain occasions multiple groups can exist simultaneously. Our data show that 
one or more QPOs are present for 78.3\% of the 361 hours of observation time. More specifically, one QPO can be observed for 41\% of the time, while two or more QPOs are present  for 37.3\%
of the time. No QPOs are present for 21.6\% of the 15 days of observations.

\begin{figure}
\begin{subfigure}{}
        \vspace{-0.2\baselineskip}
    \includegraphics[width=0.5\columnwidth]{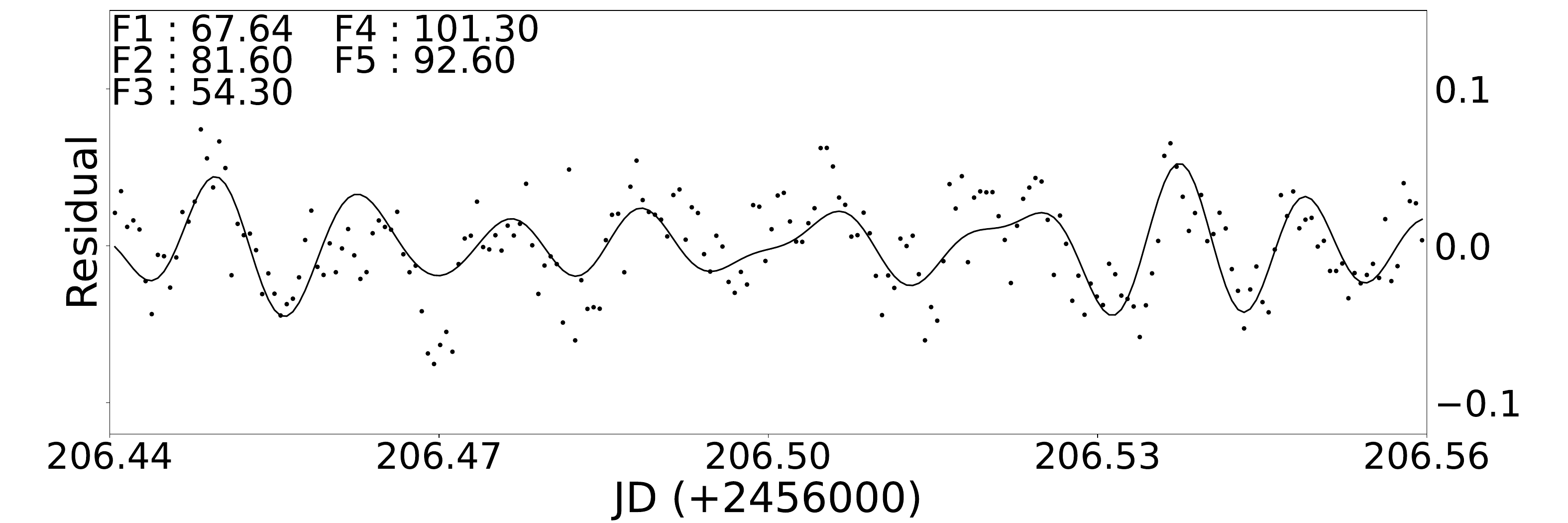}
    \end{subfigure}
    \vspace{-0.3\baselineskip}
\begin{subfigure}{}
    \vspace{-0.2\baselineskip}
        \includegraphics[width=0.5\columnwidth]{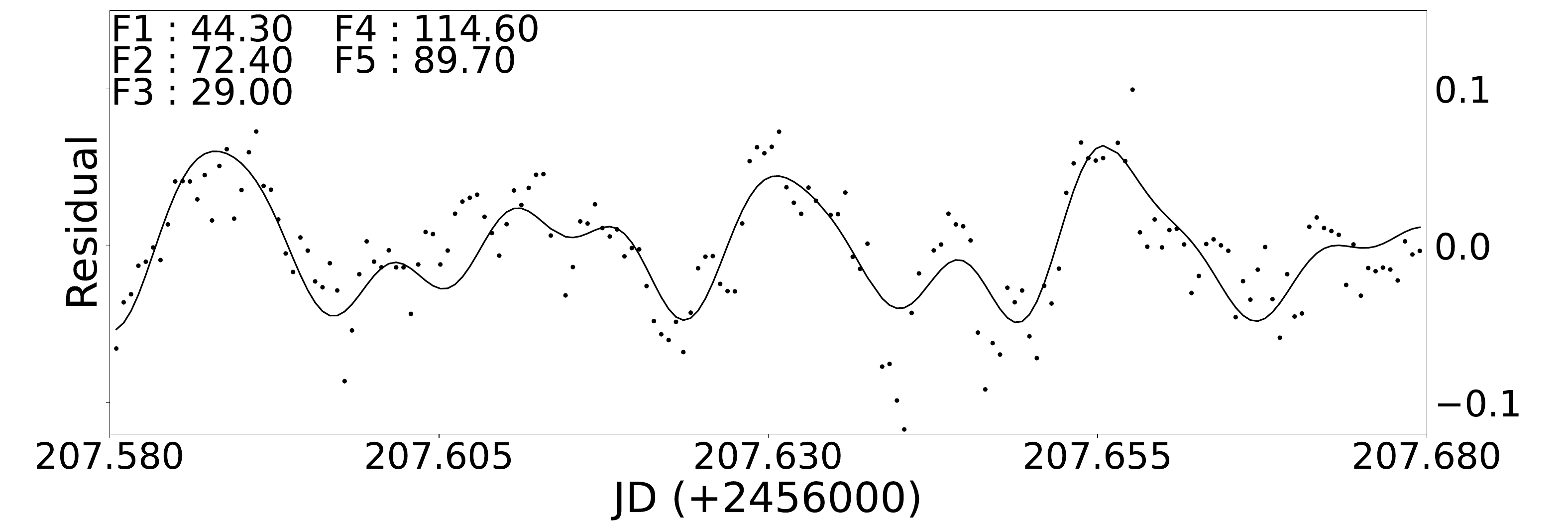}
    \end{subfigure}
    \vspace{-0.3\baselineskip}    
\begin{subfigure}{}
        \includegraphics[width=0.5\columnwidth]{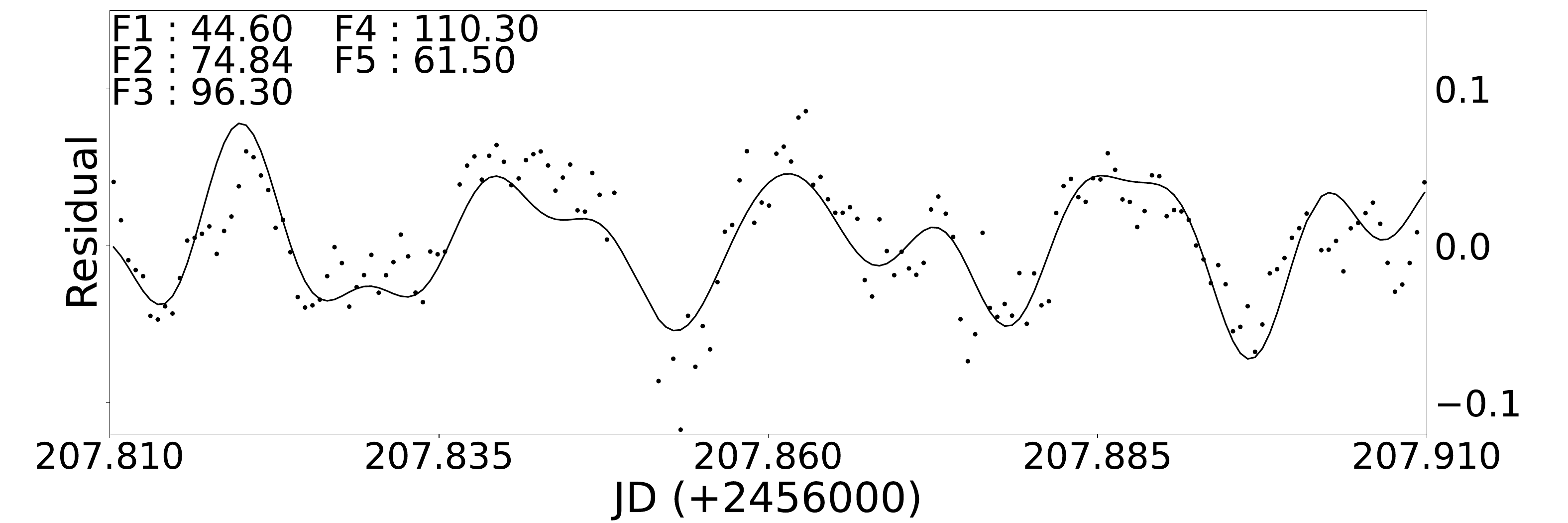}
    \end{subfigure}
    \vspace{-0.3\baselineskip}
\begin{subfigure}{}
        \includegraphics[width=0.5\columnwidth]{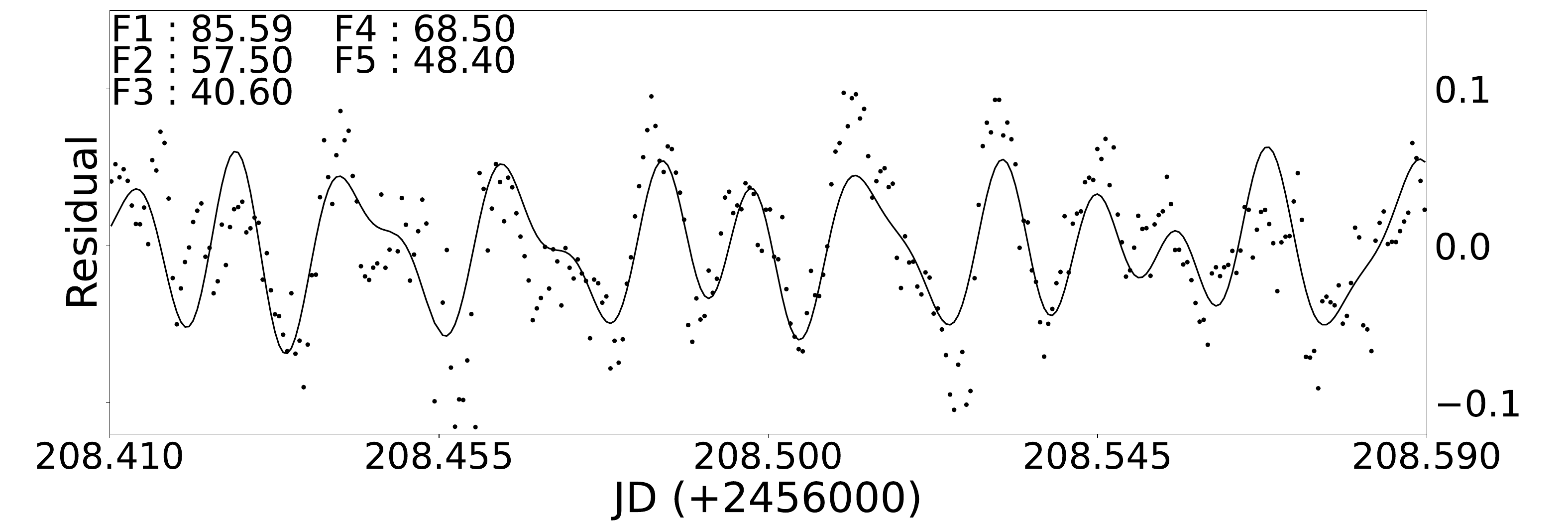}
    \end{subfigure}
    \vspace{-0.3\baselineskip}
\begin{subfigure}{}
        \includegraphics[width=0.5\columnwidth]{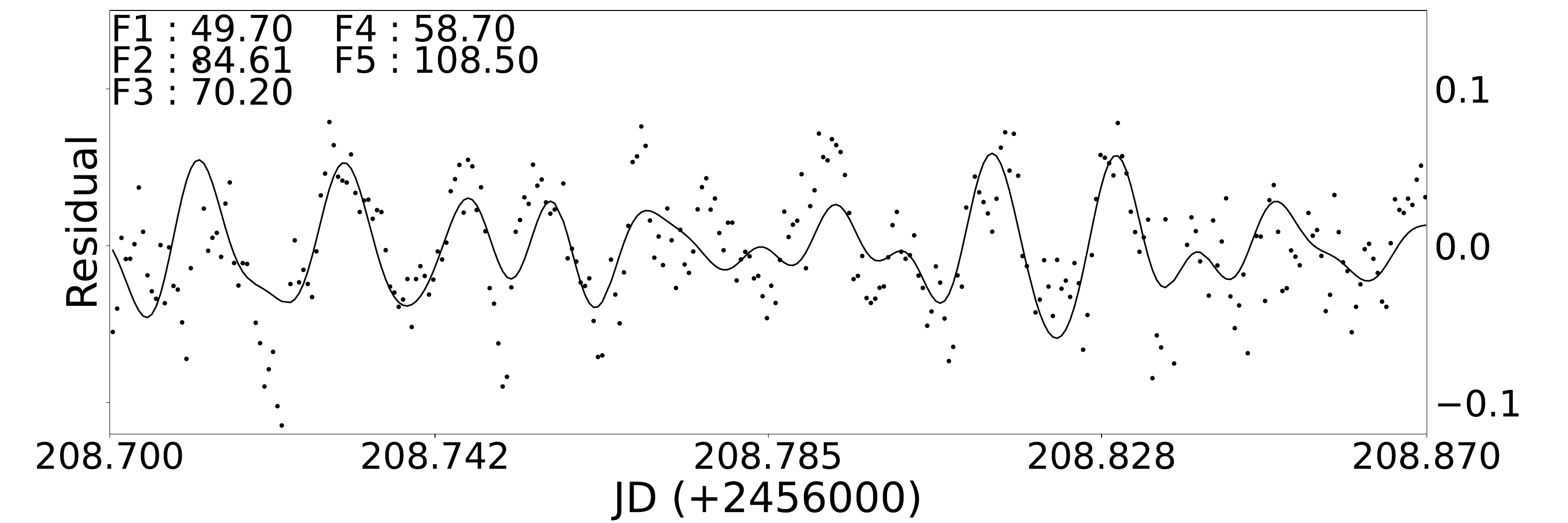}
    \end{subfigure}
    \vspace{-0.2\baselineskip}
\begin{subfigure}{}
        \includegraphics[width=0.5\columnwidth]{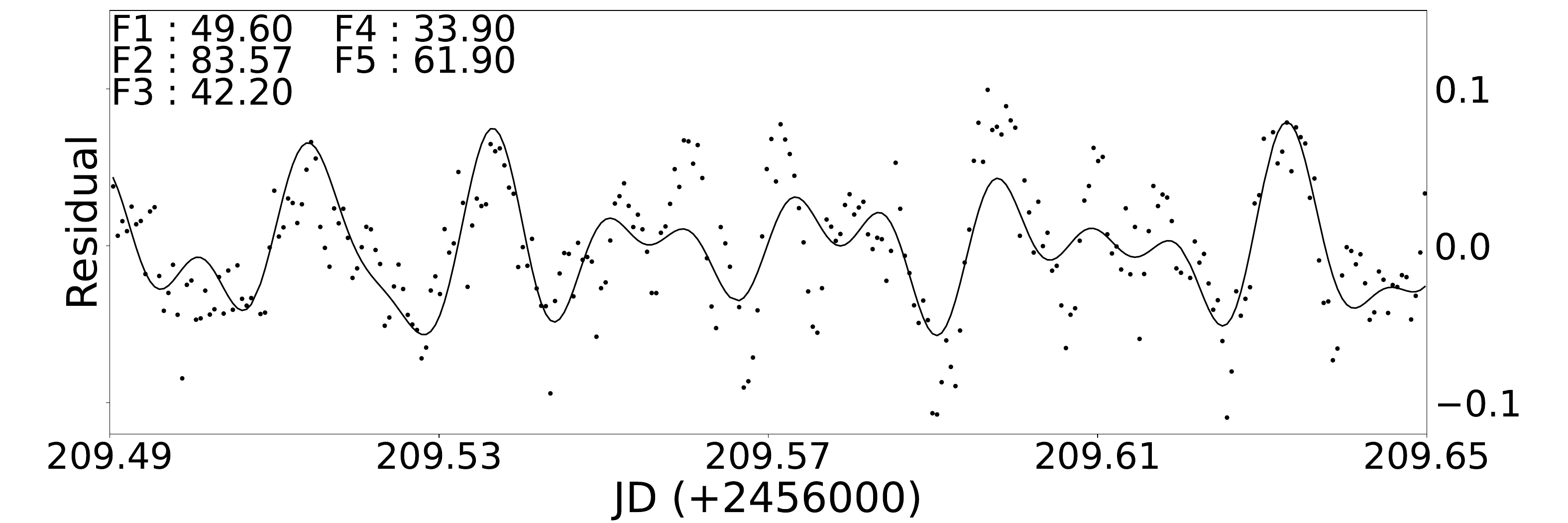}
    \end{subfigure}
    \vspace{-0.2\baselineskip}
\begin{subfigure}{}
        \includegraphics[width=0.5\columnwidth]{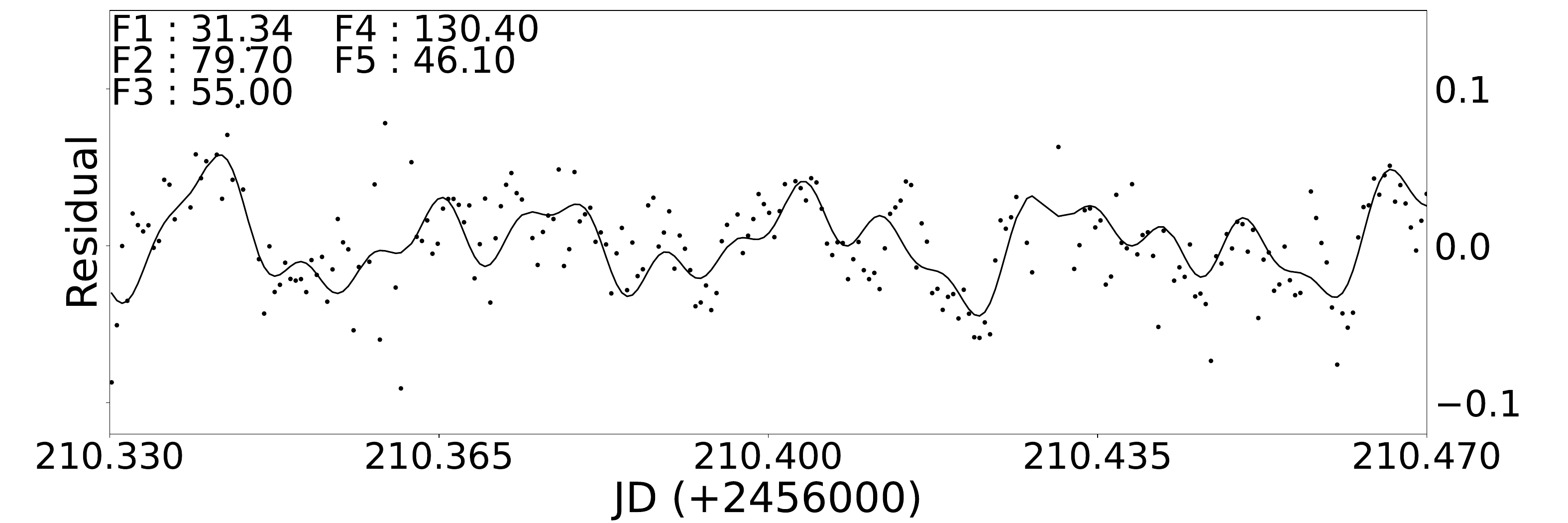}
    \end{subfigure}
    \vspace{-0.2\baselineskip}   
\begin{subfigure}{}
        \includegraphics[width=0.5\columnwidth]{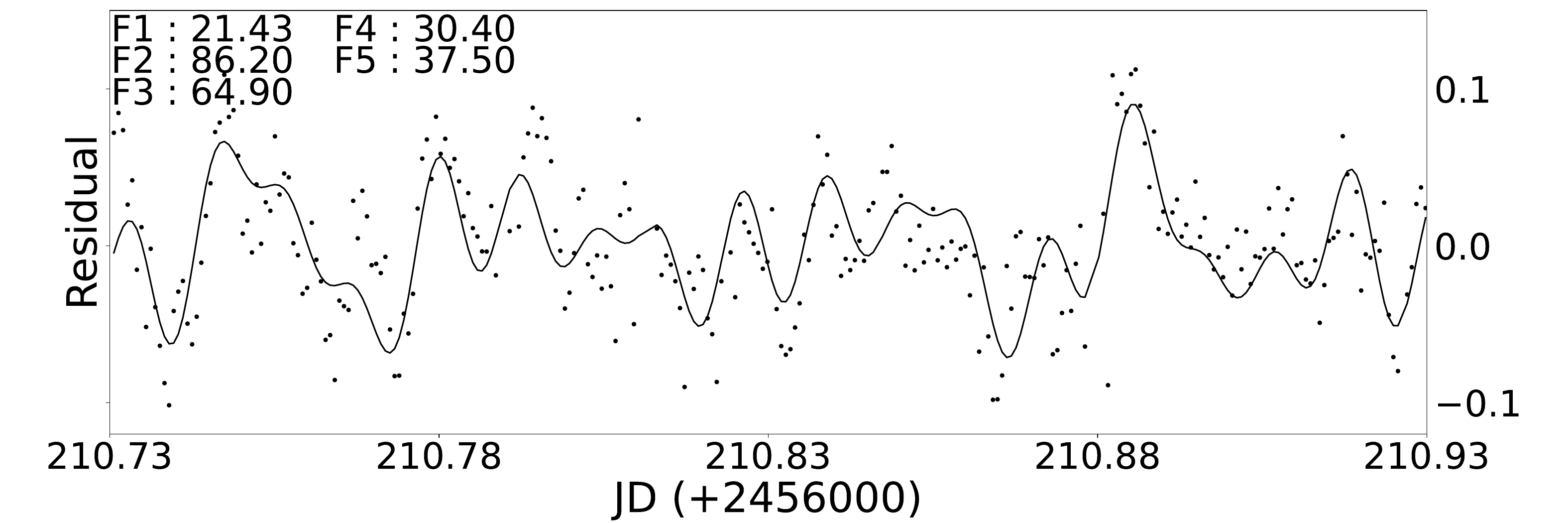}
    \end{subfigure}
    \vspace{-0.2\baselineskip}
\begin{subfigure}{}
        \includegraphics[width=0.5\columnwidth]{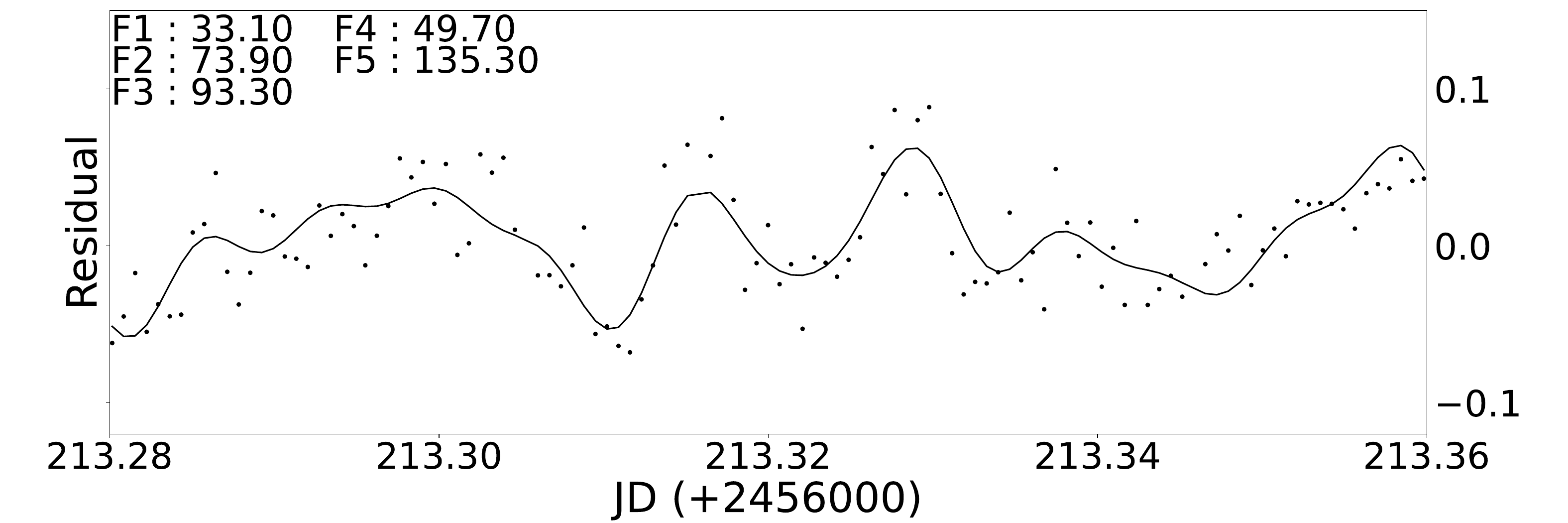}
    \end{subfigure}
    \vspace{-0.2\baselineskip}
\begin{subfigure}{}
        \includegraphics[width=0.5\columnwidth]{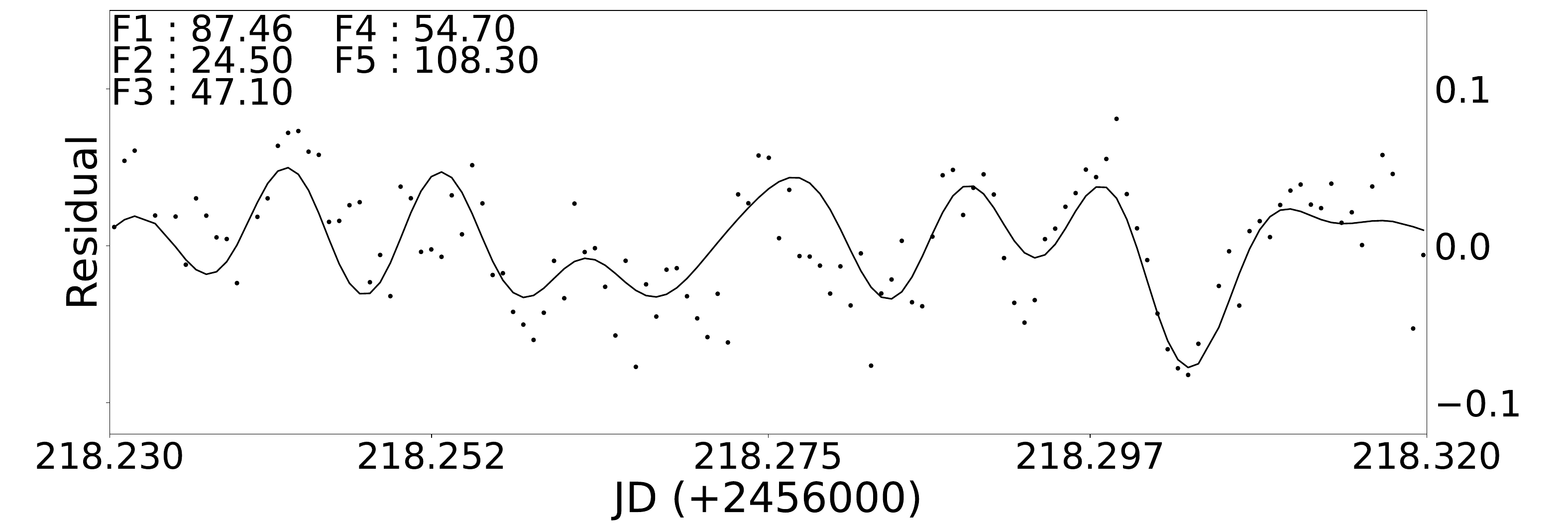}
    \end{subfigure}
    \vspace{-0.2\baselineskip}    
\begin{subfigure}{}
        \includegraphics[width=0.5\columnwidth]{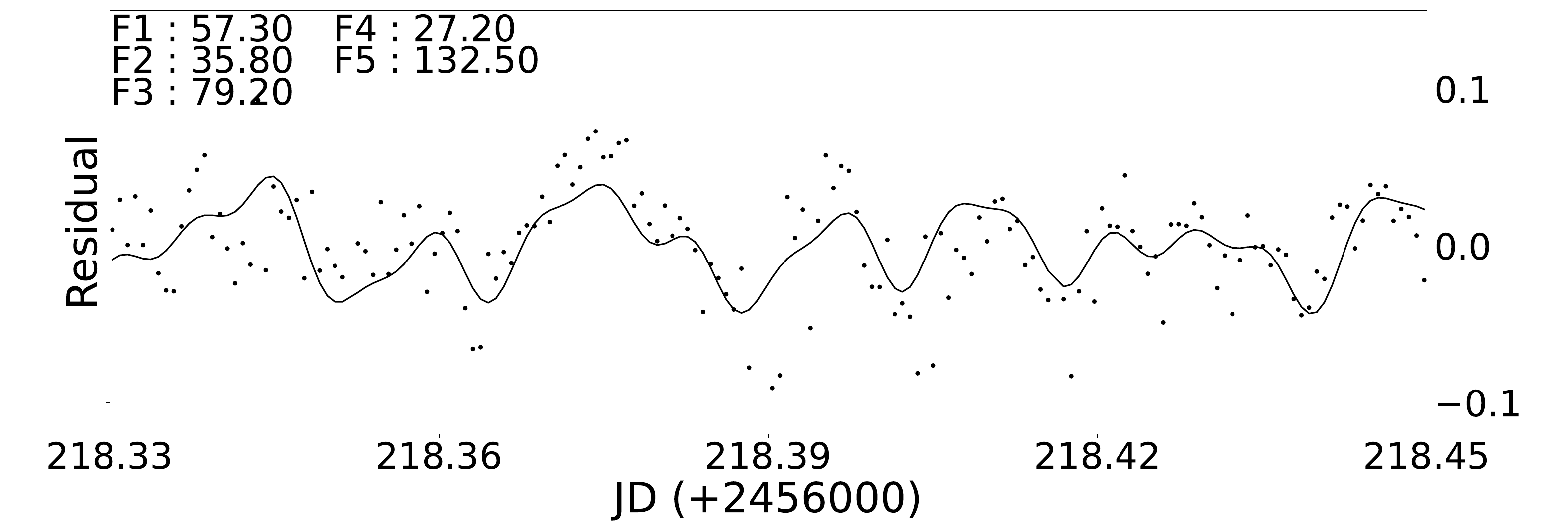}
    \end{subfigure}
    \vspace{-0.2\baselineskip}
\begin{subfigure}{}
        \includegraphics[width=0.5\columnwidth]{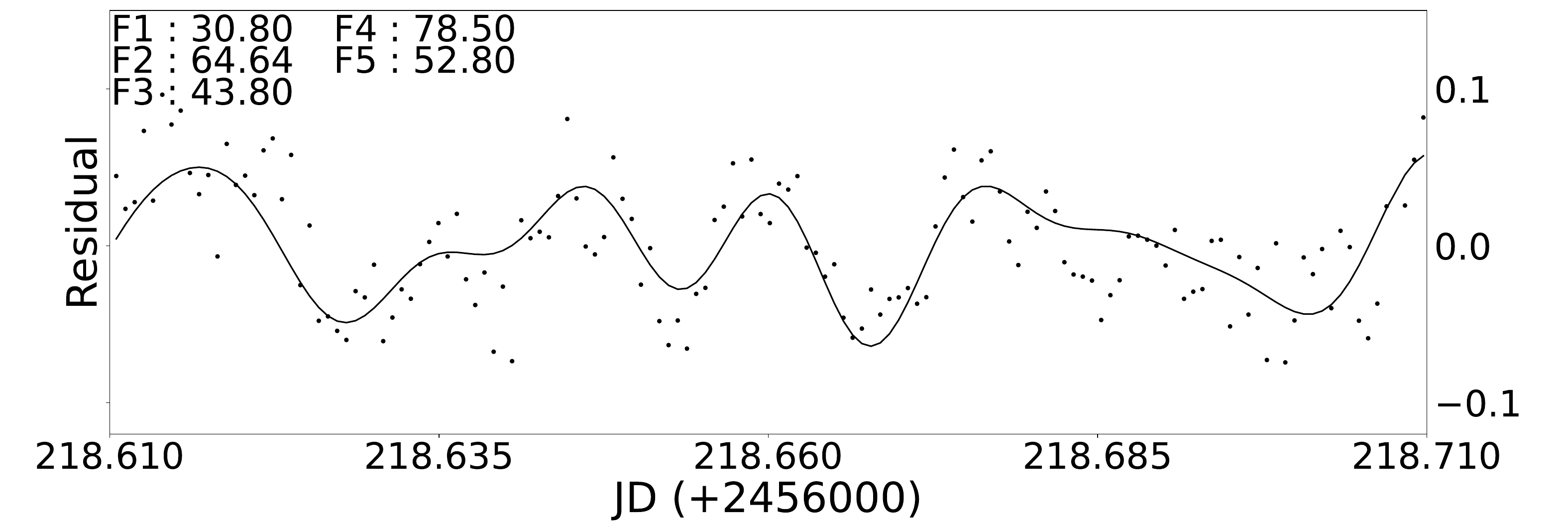}
    \end{subfigure}
    \vspace{-0.2\baselineskip}
\begin{subfigure}{}
        \includegraphics[width=0.5\columnwidth]{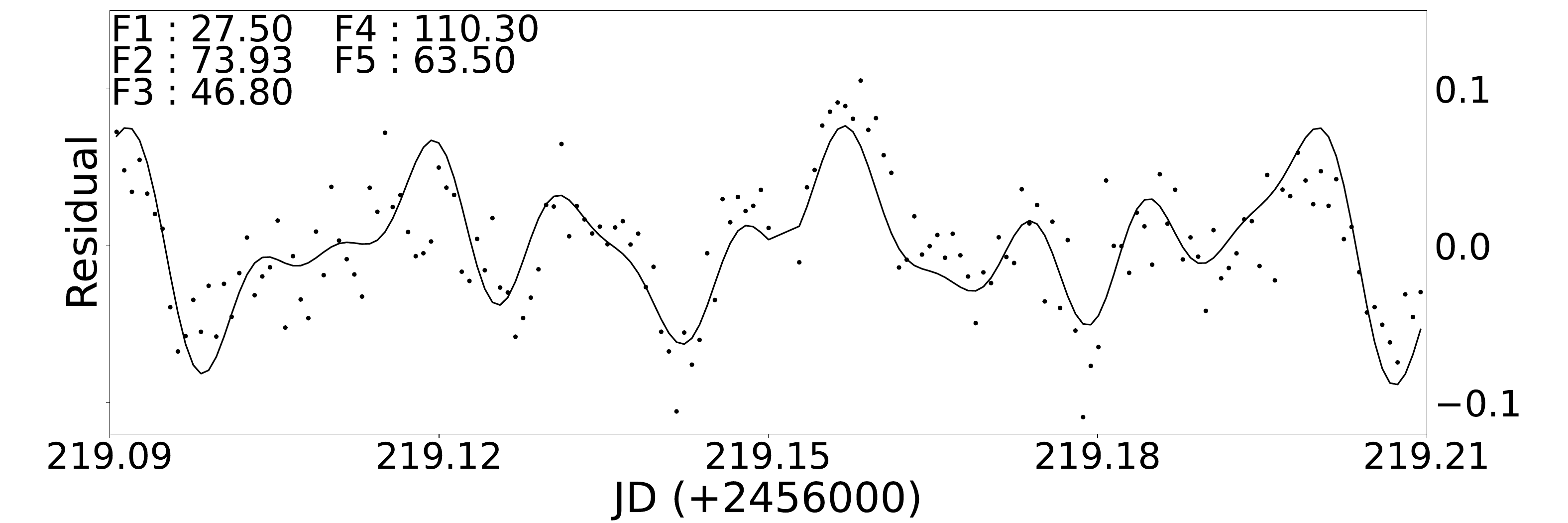}
    \end{subfigure}
    \vspace{-0.2\baselineskip}
\begin{subfigure}{}
        \includegraphics[width=0.5\columnwidth]{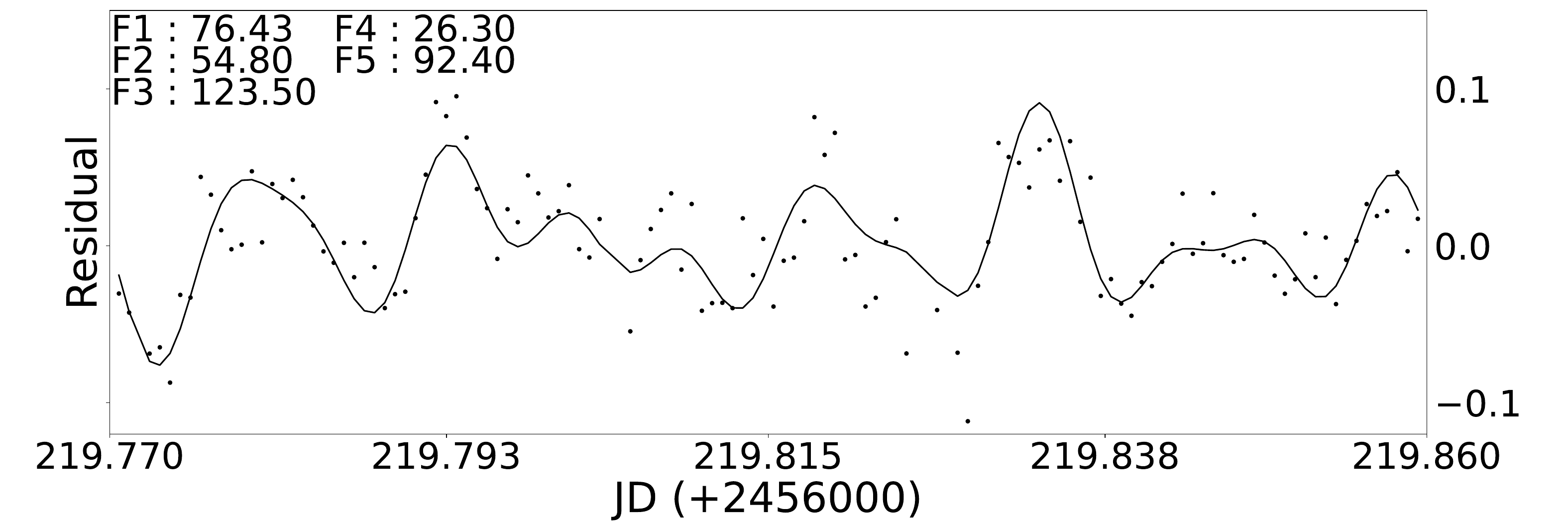}
    \end{subfigure}
    \vspace{-0.2\baselineskip}
\caption{Multiple time intervals from TT Ari’s light curve, and a sinusoidal curve overlay. Each panel shows a different frequency group. The curve is made using a combination of five significant frequencies extracted from the power-spectrum for that segment. Here, F1 corresponds to the most relevant frequency in the power spectrum, with each following frequency having diminished impact on the curve.
}
\label{fig:figure4}
\end{figure}

In Fig. \ref{fig:figure4} we show example light curves, together with the function made from the frequencies of the power-spectrum analysis. Each panel corresponds to a single frequency group using a combination of five significant frequencies from that time interval. This procedure is necessary in order to reproduce the observed behaviour. This is different from the presentation of frequency groups in Fig. \ref{fig:figure3}, where we only use a single frequency with the most relevance for each group. It is important to remember that flickering is always superimposed on the QPO light curve (see sect. \ref{sec:flick}).

\begin{figure}
        \includegraphics[width=\columnwidth]{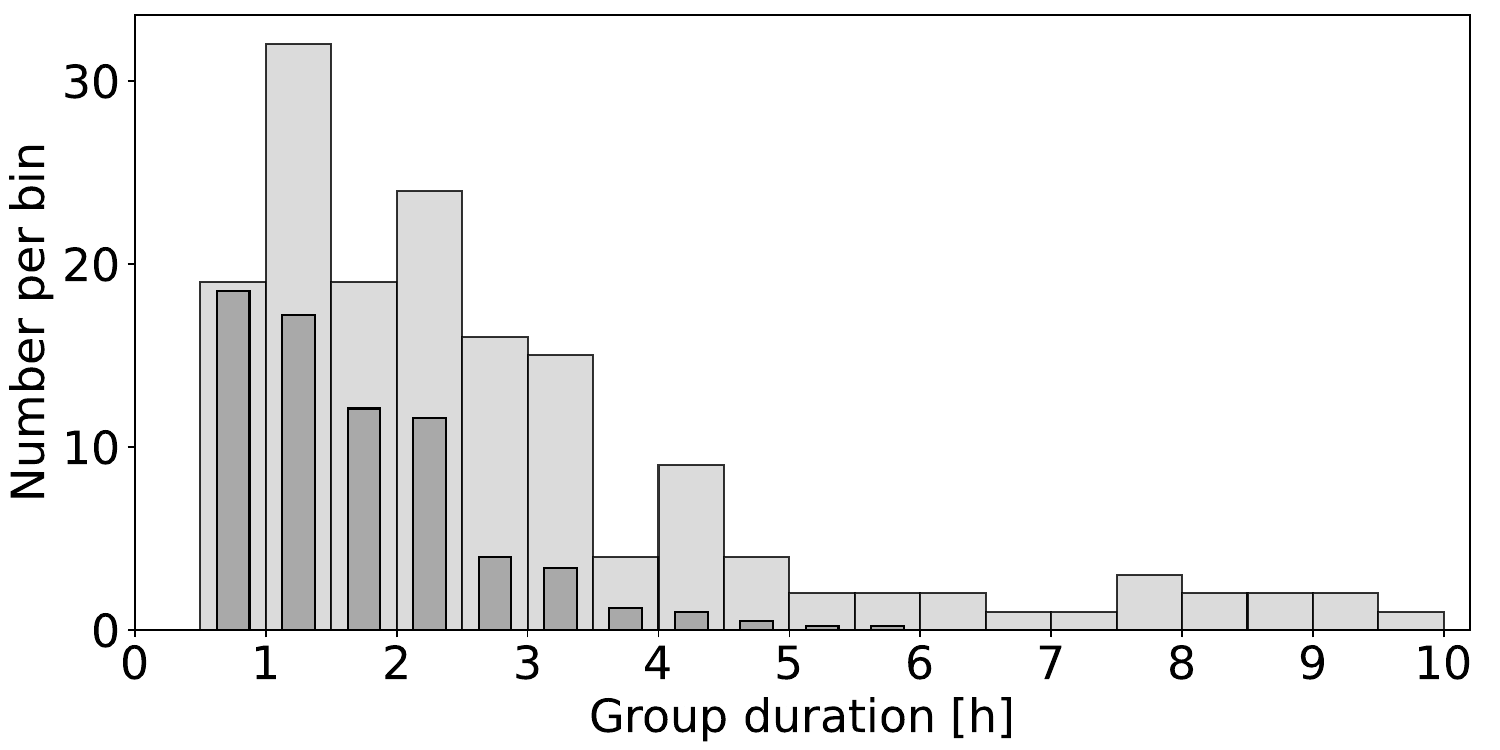}
    \caption{Group duration histograms for $P \ge 0.10$. Light grey: Duration histogram for the 160 groups of the residual data. The average duration is 2.8 hours, with few groups lasting more than 7.5 hours. Dark grey: Duration histogram for randomised data. Random data create fewer and shorter groups.
    }
    \label{fig:figure5}
\end{figure}

In Fig. \ref{fig:figure5} we can observe the duration of the groups for $P \ge 0.10$. The light grey histogram for the filtered data shows two peaks around one and two hours, respectively. It is worth considering that the minimum threshold for duration is 0.7 hours, explaining the sharp cut seen at lower values. The number of groups drops quickly after 4.5 hours, but some are still found with a duration of 10 hours.
The thin dark grey histogram shows the duration of frequency groups made from randomised data. These groups are fewer and shorter than those from the original data, which suggests that QPOs are different from stochastic processes. (see sect \ref{sec:random_data_findings}). The fact that the light grey and thin dark grey histograms have a similar shape is probably related to the definition of frequency groups: Longer groups are simply rarer and scarcer than shorter ones, regardless of the type of light curve.

Figure \ref{fig:figure6} displays the number of completed cycles ($N_{c}$) for $P \ge 0.10$. About 27\% of frequency groups have between three and four cycles, which shows that lower values are more common. The average of all groups is 7.0 cycles. The thin dark grey histogram shows the number of cycles completed by frequency groups made from randomised data. Comparing both histograms, they have a similar shape; groups with more cycles are rarer regardless of the type of light curve. As in Fig. \ref{fig:figure5}, randomised data create fewer groups overall.

\subsection{Frequency ranges of the QPOs}
\label{sec:frequency ranges}

Table \ref{tab:table2} summarises the results of all groups. It  shows the statistics of the quantities already discussed in this section. The table focuses on the comparison between two data sets; groups obtained with a normal filter, and those obtained with a strict filter. The quantity that defines the filters is power (P). For example, a filter of $P \ge 0.10$ means that only frequencies with a power greater than 0.10 in the Lomb-Scargle power spectrum are included in the set. 

For $P \ge 0.15$, $f_{\Delta}$ is more negative and $f_{A}$ and $N_{c}$ are smaller when compared with the cases of $P \ge 0.10$. This is expected, as using only frequencies with a higher power generates fewer, but more relevant groups. The table also shows $\overline{f}$ (d$^{-1}$), which is the mean frequency of each group found. For this quantity, the frequencies in the $P \ge 0.15$ column show less dispersion. The peaks and preferred range of the frequencies of the groups can be seen in Fig. \ref{fig:figure7}.

The histograms of Fig. \ref{fig:figure7} show all the individual frequencies that constitute frequency groups, from different filters. The light grey histogram corresponds to $P \ge 0.10$. There are fewer cases below a frequency of 25 d$^{-1}$ and over 100 d$^{-1}$. There is a gap in the middle between 50 and 65 d$^{-1}$ and peaks occur at frequencies of 42.5 and 77.5 d$^{-1}$. These peaks do not appear to be caused by harmonics, as they have different shapes and 42.5 is not half of 77.5. The grey histogram corresponds to frequencies for the filter $P \ge 0.15$. The general shape of the histogram is similar, with fewer cases below a frequency of 30 d$^{-1}$ and over 90 d$^{-1}$; there is a gap in the middle between 55 and 75 d$^{-1}$, and peaks appear at frequencies of 42.5 and 80 d$^{-1}$. As the filter is stricter here, there are fewer cases among all frequencies, explaining the lower count in the histogram. Meanwhile, the thin histograms correspond to the frequencies of the randomised data. The shuffling process was repeated 100 times and the average number of frequencies per bin was calculated. This average is the reason why both thin histograms form Gaussian curves. In contrast, random distributions exhibit peaks at random frequencies, as can be seen in Fig. \ref{fig:figure11}. Thin histograms have fewer frequencies, but the frequencies themselves are found at higher values: the averages are $\overline{f}=83$ and $\overline{f}=68.4$ for random and original data, respectively. Table \ref{table3_new} shows this and other parameters. The thin light grey histogram corresponds to $P \ge 0.10$, and has more cases than the thin dark-grey histogram, which uses $P \ge 0.15$. Considering all four histograms in Fig. \ref{fig:figure7}, our data suggest that QPOs are more common at certain frequencies. This trend continues across filters and is qualitatively different from random processes. Furthermore, the frequencies found are part of groups, which means that QPOs have a permanence in time and their evolution can be tracked.

\subsection{QPO evolution}
Figure \ref{fig:figure8} corresponds to the histogram of the frequency delta ($f_{\Delta}$) and frequency amplitude ($f_{A}$).
These quantities have the following implications: If $f_{\Delta} > 0$, the group's mean frequency increases over time, and if $f_{\Delta}=0$, the group either has a constant frequency, or it changes and returns to its original frequency. As for the frequency amplitude, if $f_{A}=0,$ then the group remains at a single frequency throughout its entire duration. The bottom histogram shows that this is uncommon; most groups change with time. On the top panel of Fig. \ref{fig:figure8}, the $f_{\Delta}$ histogram has a central peak around $-1.0$\,d$^{-1}$, and a second, smaller peak at $-7.0$\,d$^{-1}$. There is a similar number of positive and negative cases; the average is $\overline{f}_{\Delta}=-0.24$ \,d$^{-1}$. The negative value means that it is slightly more common for groups to evolve towards lower frequencies than towards higher ones. The bottom panel of Fig. \ref{fig:figure8} shows that $f_{A}$ is not centred around zero. Instead it peaks at around $7.0$\, d$^{-1}$ with an average of $9.9$ \,d$^{-1}$. This implies that a frequency can continually shift up or down by $9.9$ \,d$^{-1}$ on average; this transition is sufficiently smooth that the frequency is still recognised as being part of the same frequency group. These results imply that most frequencies evolve slowly and very few remain constant. The number of cases decreases quickly after 18 d$^{-1}$.

\clearpage
\newpage
\onecolumn

\clearpage
\twocolumn

\begin{table*}
\caption{Average parameters of frequency groups.}             
\label{tab:table2}      
\centering
\resizebox{\textwidth}{!}{%
\begin{tabular}{@{}lrrrrrrrrrr@{}}     
\hline
\noalign{\smallskip}
& \multicolumn{2}{c}{$\overline{f}$ (d$^{-1}$)\,\tablefoottext{a}} & \multicolumn{2}{c}{Duration (d)\,\tablefoottext{b}} & \multicolumn{2}{c}{$f_{\Delta}$\,(d$^{-1}$)\,\tablefoottext{c}} & \multicolumn{2}{c}{$f_{A}$ (d$^{-1}$)\,\tablefoottext{d}} & \multicolumn{2}{c}{$N_{c}$\,\tablefoottext{e}} \\ 
\noalign{\smallskip}
\hline
\noalign{\smallskip}
 & P > 0.10 & P > 0.15 & P > 0.10 & P > 0.15 & P > 0.10 & P > 0.15 & P > 0.10 & P > 0.15 & P > 0.10 & P > 0.15 \\
\noalign{\smallskip}
\hline    
\noalign{\smallskip}
Peaks\,\tablefoottext{f} & 42.5 and  77.5 & 42.5  and 80.0 & ... & ... & ... & ... & ... & ... & ... & ... \\
Preferred range\,\tablefoottext{g} & 27--98 & 28--91 & ... & ... & ... & ... & ... & ... & ... & ... \\
Minimum & 20.01 & 21.39 & 0.03 & 0.03 & -29.6 & -22.0 & 1.0 & 0.6 & 3.0 & 3.0 \\
Maximum & 126.11 & 114.66 & 0.40 & 0.37 & 29.0& 14.3 & 33.1 & 24.3 & 29.0 & 18.0 \\
Mean & 68.45 & 66.14& 0.12& 0.11 & -0.2 & -0.7 & 9.9 & 9.4 & 7.0& 6.5 \\
Standard deviation & 24.640 & 22.088 & 0.083 & 0.070 & 8.62 & 8.07 & 6.32 & 5.25 & 4.42 & 3.70 \\
Standard error of Mean & 1.945 & 2.469 & 0.006 & 0.008 & 0.68 & 0.90 & 0.49 & 0.59 & 0.35 & 0.41 \\ 
\hline 
\end{tabular}%
}
\tablefoot{
For each quantity, there are two columns, each representing a filter with different power ($P$): a column for frequencies with  $P \ge 0.10$ in the power spectrum, and another one for $P \ge 0.15$. $f_{\Delta}$, $f_{A}$, and  $N_{c}$ are explained in sect. \ref{sec:frequency groups}.\\
\tablefoottext{a}{Mean frequency of each group.}
\tablefoottext{b}{Duration of each group.}
\tablefoottext{c}{Frequency delta.}
\tablefoottext{d}{Frequency amplitude.}
\tablefoottext{e}{Number of completed cycles.}
\tablefoottext{f}{Frequencies where most groups are found.}
\tablefoottext{g}{Frequency range that encompasses 90\% of the data.}
}
\end{table*}

\begin{table}
\caption{Comparison of frequency groups created from synthetic (random) light curves and the original light curves.}            
\label{table3_new} 
\resizebox{\columnwidth}{!}{%
\begin{tabular}{@{}lllll@{}}                             
\noalign{\smallskip}
\hline   
& \multicolumn{2}{c}{Synthetic}& \multicolumn{2}{c}{Original} \\
\noalign{\smallskip}
\hline   
                  & $P\ge0.10$ & $P\ge0.15$ & $P\ge0.10$ & $P\ge0.15$ \\
\noalign{\smallskip}
\hline      
N° groups\,\,\tablefootmark{a}          & 70   & 14   & 160       & 80        \\
N° $f$\,\,\tablefootmark{b}             &596   & 112  & 2019      & 959       \\
$\overline{f}$\,\,\tablefootmark{c}     & 83   & 82.4 & 68.4      & 66.1      \\
Mean duration\,\tablefootmark{d}        & 1.7  & 1.5  & 2.8       & 2.6       \\
$\overline{N_{c}}$\,\,\tablefootmark{e} & 5.9  & 5.2  & 7.0       & 6.5       \\
\hline                                
\end{tabular}
}
\tablefoot{
For all the parameters related to random data, the shuffling process was repeated 100 times and the average was calculated.\\
\tablefoottext{a}{Number of groups found.}
\tablefoottext{b}{Number of frequencies forming the groups.}
\tablefoottext{c}{Mean frequency (d$^{-1}$).}
\tablefoottext{d}{Mean duration (hours).}
\tablefoottext{e}{Mean number of cycles.}
}
\end{table}
%

\begin{figure}
        \includegraphics[width=\columnwidth]{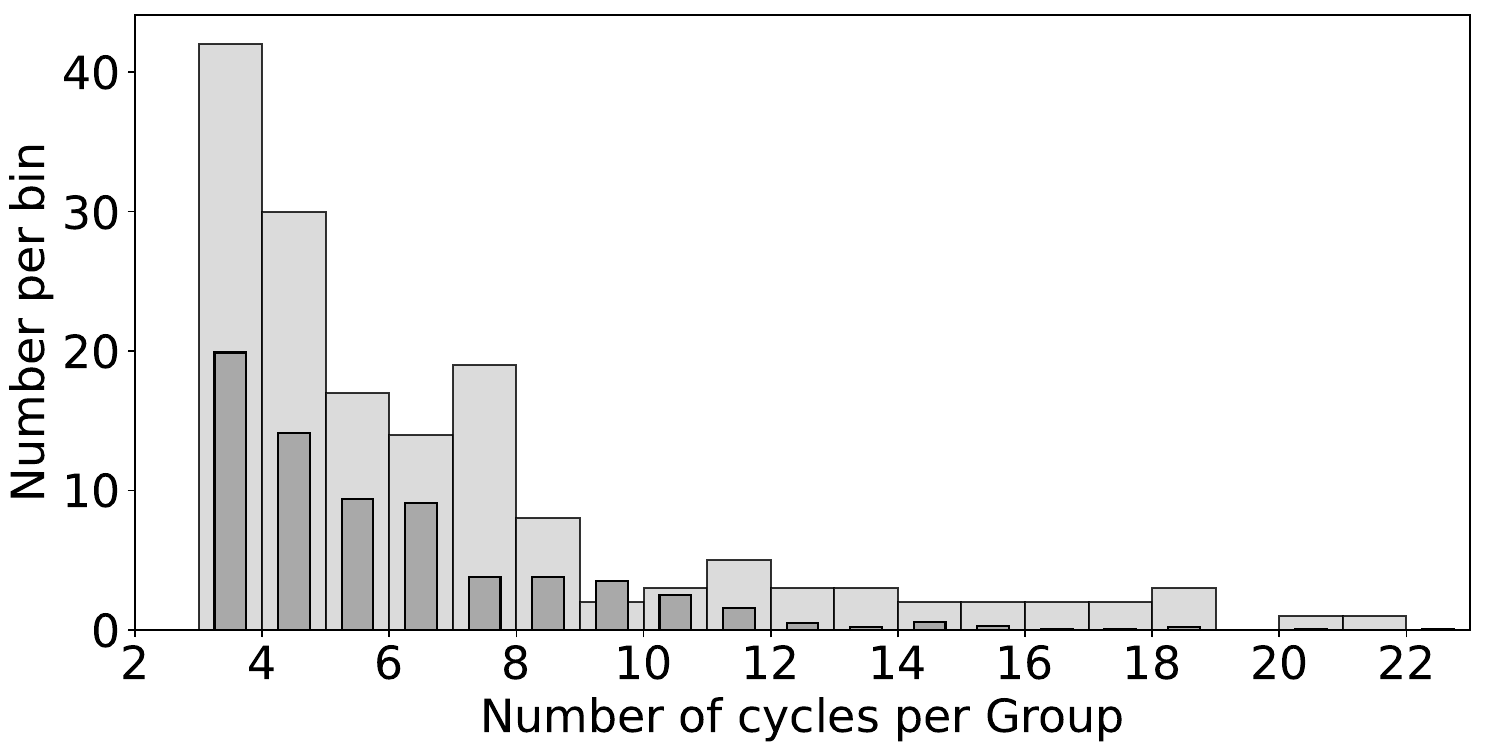}
    \caption{Number of completed cycles $N_{c}$ for $P \ge 0.10$. Light grey: Histogram for the 160 groups of the residual data. On average, a group can complete seven cycles before shifting to other frequencies. Thin dark grey: Duration histogram for randomised data. Groups made from random data complete fewer cycles on average.  
}
    \label{fig:figure6}
\end{figure}

\subsection{Correlations}

All well-defined superhump maxima in our light curve obey the ephemeris $\mathrm{HJD} = 2\,456\,206.0575(17) + 0.132897(26)E$, as determined by a linear fit. This period is slightly smaller than that of $P_{s} = 0.13309(8)$ days derived  for TT Ari from earlier MOST data by \cite{Vogt_2013}, underlining the typical stochastic variability of the superhump period of TT Ari. When comparing the phase of the above superhump ephemeris with parameters such as the mean frequency, group duration, number of cycles, mean power, frequency delta, and the frequency amplitude, we find no significant correlation, because the corresponding Pearson coefficients are within the range of $-0.12<r<0.15$.
 
The top panel of Fig. \ref{fig:figure9} presents the correlation between the mean frequency (d$^{-1}$) and mean power for the frequency groups. As expected, there are a greater number of points in the most common frequencies (in the 30--90 d$^{-1}$ range). The strength of the correlation is moderate, with a Pearson coefficient (r) of $r= -0.39$ and a p-value (p) of $p=3.09*10^{-7}$. The slope found is probably the result of high-frequency groups not having high values in the power spectrum. In the middle panel of Fig. \ref{fig:figure9}, there is a moderate correlation ($r= -0.48$, $p=1.53*10^{-10}$) between the mean frequency (d$^{-1}$) of each group and their duration in JD. However, this correlation might by caused by an area without data in the bottom left corner. As cycles are defined as the product of duration and frequency, this means that there are no groups with both low frequency and low duration, as they would not reach three completed cycles, the minimum threshold. The groups with longer duration are located near the most common frequencies (42.5 and 77 d$^{-1}$). In the bottom panel of Fig. \ref{fig:figure9}, we compare mean power and duration (JD). The strength of the correlation is high ($r= 0.68$, {$p=5.96*10^{-23}$). It seems that frequencies with high power create longer groups.

\begin{figure}
        \includegraphics[width=\columnwidth]{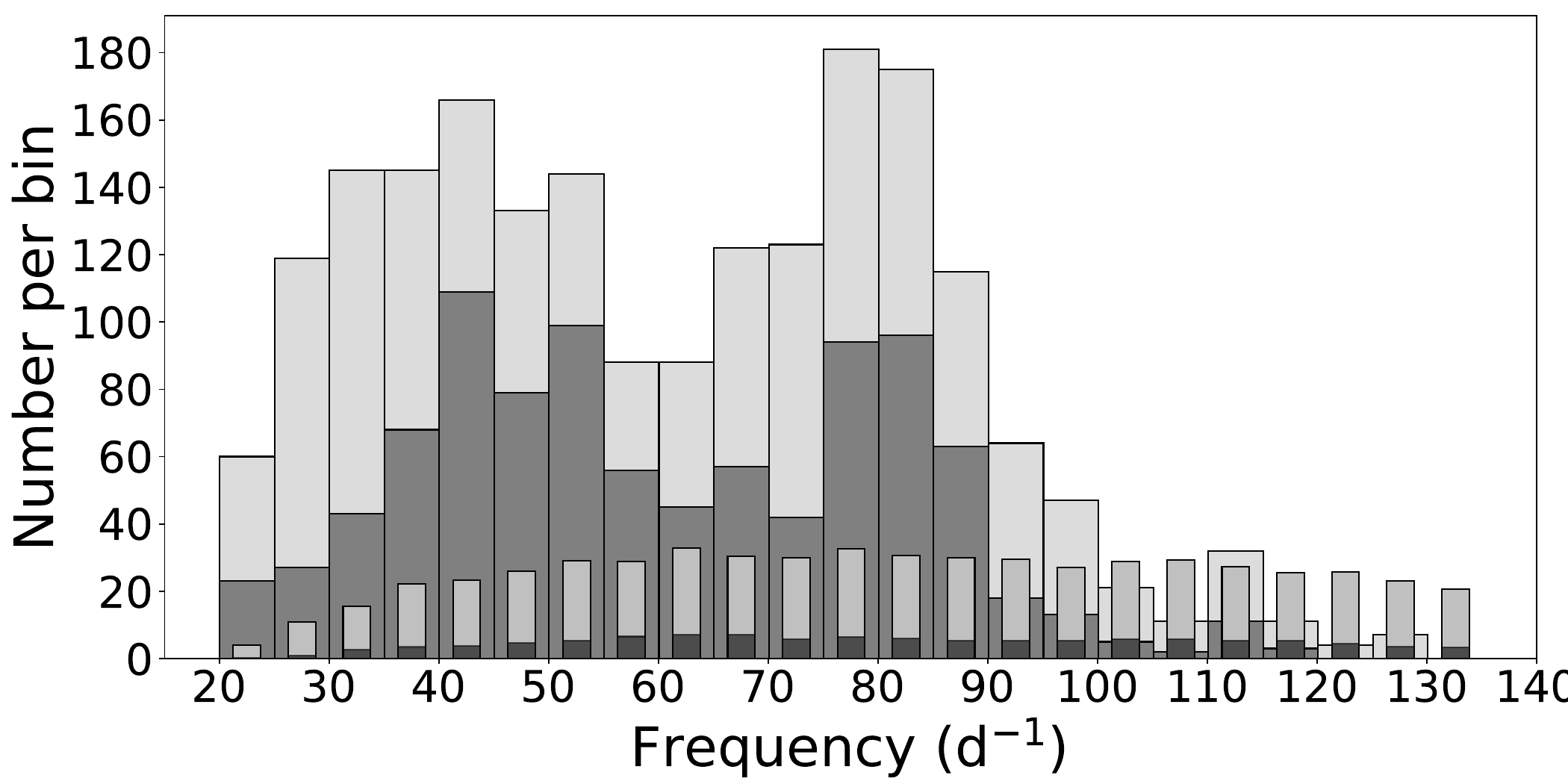}
    \caption{Comparison of different frequency histograms for all the frequencies that constitute the frequency groups. The light grey histogram shows the frequencies of the filtered data, with a $P \ge 0.10$ from the original light curve. Similarly, the dark grey histogram represents the frequencies of the filtered data but with a higher threshold for power spectrum, $P \ge 0.15$. The thin grey histogram shows the frequencies of randomised light curves with $P \ge 0.10$ in the power spectrum. The thin dark grey histogram is the same as the previous one, but with $P \ge 0.15$.
}
    \label{fig:figure7}
\end{figure}

 Figure \ref{fig:figure10} shows correlation plots regarding the frequency amplitude. The frequency amplitude is compared with the mean frequency (d$^{-1}$), duration (JD), number of cycles, and mean power. A moderate correlation ($r= -0.41$, $p=7.67*10^{-8}$) is found between the frequency amplitude and the mean frequency of each group (top-left). The groups with greater frequency amplitude are located near the most common frequencies (42.5 and 77 d$^{-1}$). The top-right panel shows the correlation between frequency amplitude and duration, with $r= 0.71$, $p=2.70*10^{-25}$, while the bottom-left panel shows frequency amplitude versus the number of cycles, with $r= 0.60$, $p=9.3*10^{-17}$. Both of these correlations are strong, but this is expected, because a large group amplitude and a large number of cycles require a long group. The frequency amplitude versus mean power plot in the bottom-right panel shows a moderate correlation, with $r= 0.55$ and $p=5.88*10^{-14}$. It appears that the power of a frequency group is related to its variability and frequency evolution.

\begin{figure}
\begin{subfigure}{}
        \includegraphics[width=\columnwidth]{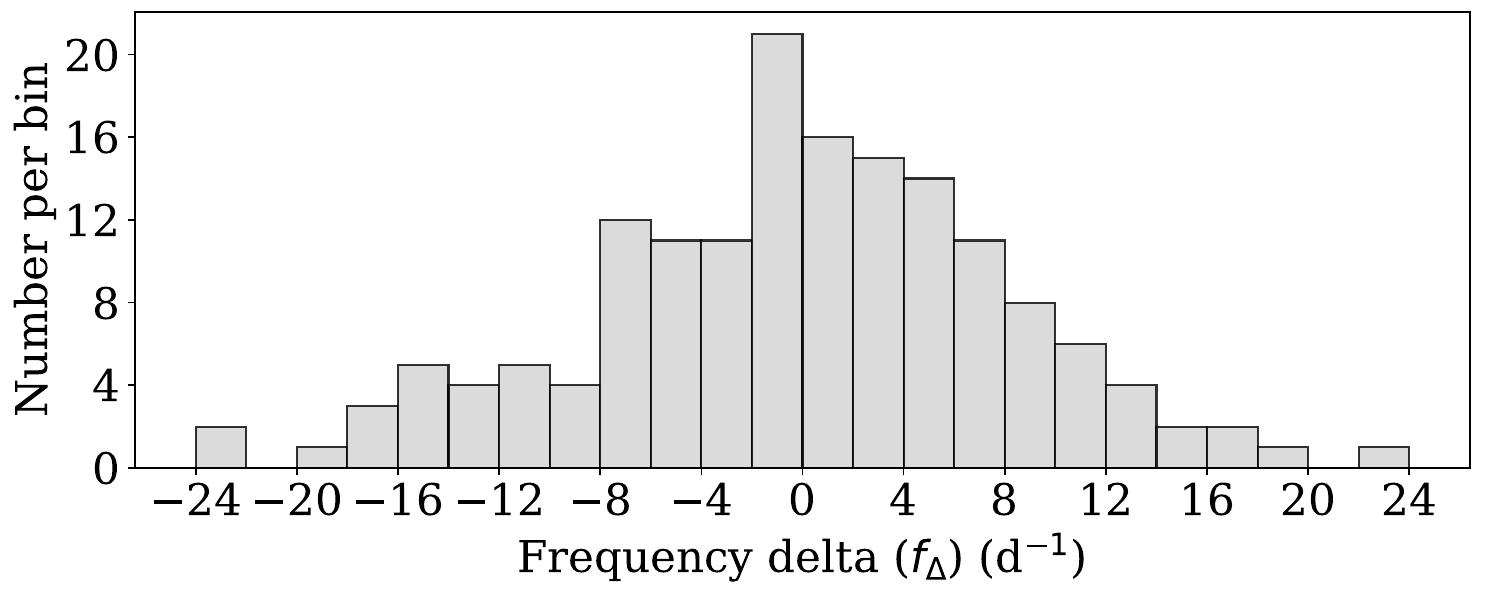}
    \end{subfigure}
    \vspace{-1.2\baselineskip}

\begin{subfigure}{}
        \includegraphics[width=\columnwidth]{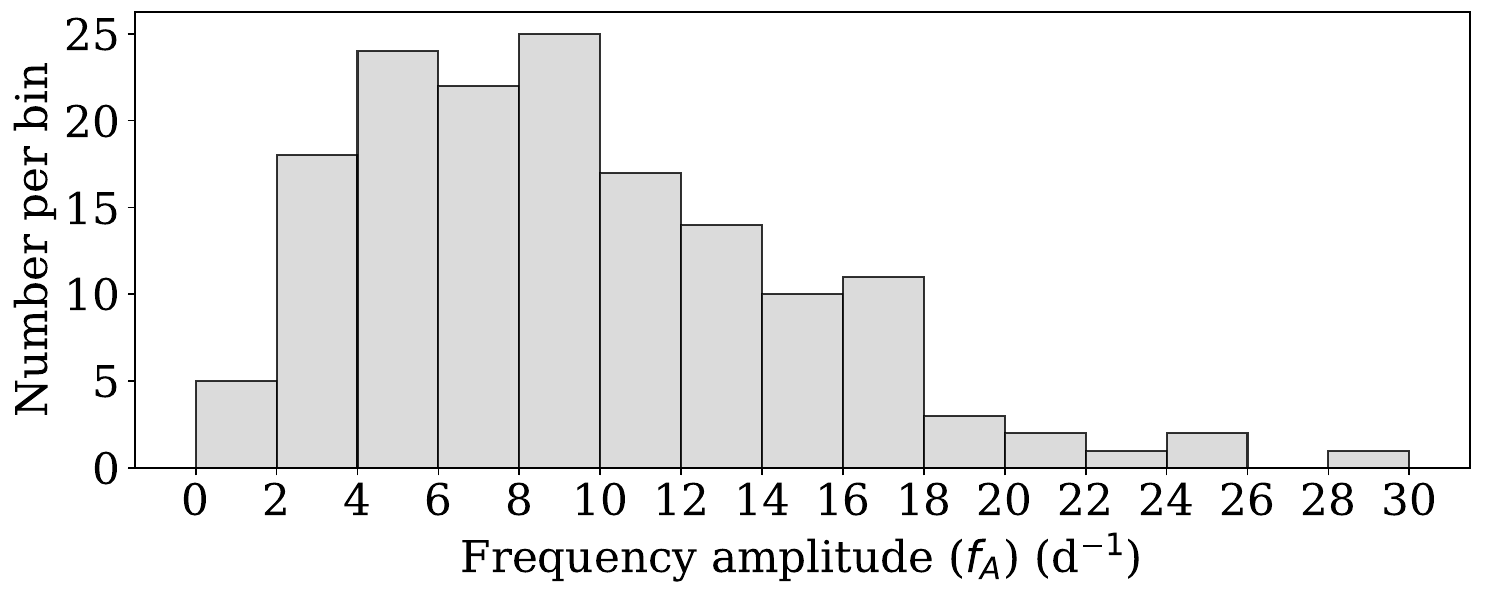}
    \end{subfigure}

\caption{
Histograms for $P \ge 0.10$. The frequency delta ($f_{\Delta}$) histogram in the top panel shows a peak near zero, which means that many groups end at the same frequency in which they started. Data points under $-22$ d$^{-1}$ were added to an underflow bar. The frequency amplitude ($f_{A}$) histogram in the bottom panel shows that most groups show changes in frequency throughout their duration. Both $f_{\Delta}$ and $f_{A}$ are described in sect. \ref{sec:frequency groups}.
}
\label{fig:figure8}
\end{figure}

\subsection{Group significance for the random data}
\label{sec:random_data_findings} 
We performed a test to see if frequency groups would appear after shuffling the data points of the light curve. The methods applied are listed in Sect. \ref{sec:random_data}.

The shuffling process was repeated 100 times. Each time, we recorded the number of frequency groups created and their relevant quantities. We then calculated the averages for these parameters, which are shown in Table \ref{table3_new}. Some examples of these iterations are presented in Fig. \ref{fig:figure11}. For $P \ge 0.10,$ we found 7009 groups, which is an average of 70 frequency groups per iteration. Compared with the results for the original light curve, the false positives could amount to a considerable 43\%. For $P \ge 0.15$ we found an average of 14 frequency groups per iteration, which is equal to 17\% of the original groups for this power threshold. Another quantity of interest is the number of frequencies that form the frequency groups. As the groups from random data are fewer and shorter, the count of frequencies is much lower, as can be seen in the thin histograms of Fig. \ref{fig:figure7}. For $P \ge 0.10$, random data contained 596 individual frequencies, which is 29.5\% of the total number of frequencies found in the original light curve, while for $P \ge 0.15,$ random data showed 112 frequencies, which is 11.7\% of the number of frequencies in the original light curve.

Based on these results, the impact of stochastic variations on our results is non-negligible. However, these random effects cannot  fully explain our findings, as the false positives are fewer, smaller, and have less recognisable features compared with the real light curve. Our view is that the results in this work are not just the consequence of random processes occurring in TT Ari, but a manifestation of some periodic process occurring in the binary system.

\section{Discussion}
Our extended data set allows a new approach to analysing QPOs. This study focuses on groups of frequencies rather than individual ones, which gives us unique insight into the variations of TT Ari. Our new method allows us to exclude frequencies with high power but low duration, which are unlikely to represent real periodicities in QPOs.

\subsection{Properties of the quasi-periodic oscillations of TT Ari} 

Our results corroborate the presence of QPOs. We confirm previous research, while improving the precision of some QPO parameters, such as duration, frequency, and occurrence. 

The periods of QPOs in TT Ari exhibit significant variability over a few hours (Fig. \ref{fig:figure7}), demonstrating preferences for specific frequency ranges. The predominant range is 25 to 100 d$^{-1}$ (16 to 57 minutes), with peaks at 43.5 and 77.5 d$^{-1}$ (18.5 and 33 minutes). This range aligns with findings from other authors (see Table \ref{tab:table1}). Earlier studies (such as \citet{Semeniuk_1987} and \citet{Hollander_1992}) were based on small samples with limited observation time. These authors found individual frequencies to describe QPOs. Considering the dynamic nature of QPOs, we decided instead to search for frequency ranges where QPOs are more common, which is the method used by \citet{Kraicheva_1999},  \citet{Andronov_1999}, and  \citet{2014AcA....64..167S}.  Still, most single frequencies from earlier works fall within the range 25 to 100 d$^{-1}$.

We also find cases where two or more QPO periods  are simultaneously present in a given segment of observations. This happened in $\sim$37\% of the observation time analysed in the present study. Similar occurrences were noted by \cite{2014AcA....64..167S}. 
Many of the segments that contain multiple QPOs show two peaks in frequency (around 43.5 and 77.5 d$^{-1}$), as seen in Figs. \ref{fig:figure3} and \ref{fig:figure7}. The latter shows a gap in the middle of the histogram, which makes the dual peaks more relevant. The second peak does not occur at double the frequency of the first one. Additionally, the shapes of the two peaks are different, and so the likelihood of one of them being a harmonic of the other is low. 
The possibility of having two peaks in frequency is also noted by \cite{Bruch_2019}, who notes that on certain observation nights the QPOs split into two `branches'.

\begin{figure}
        \includegraphics[width=\columnwidth]{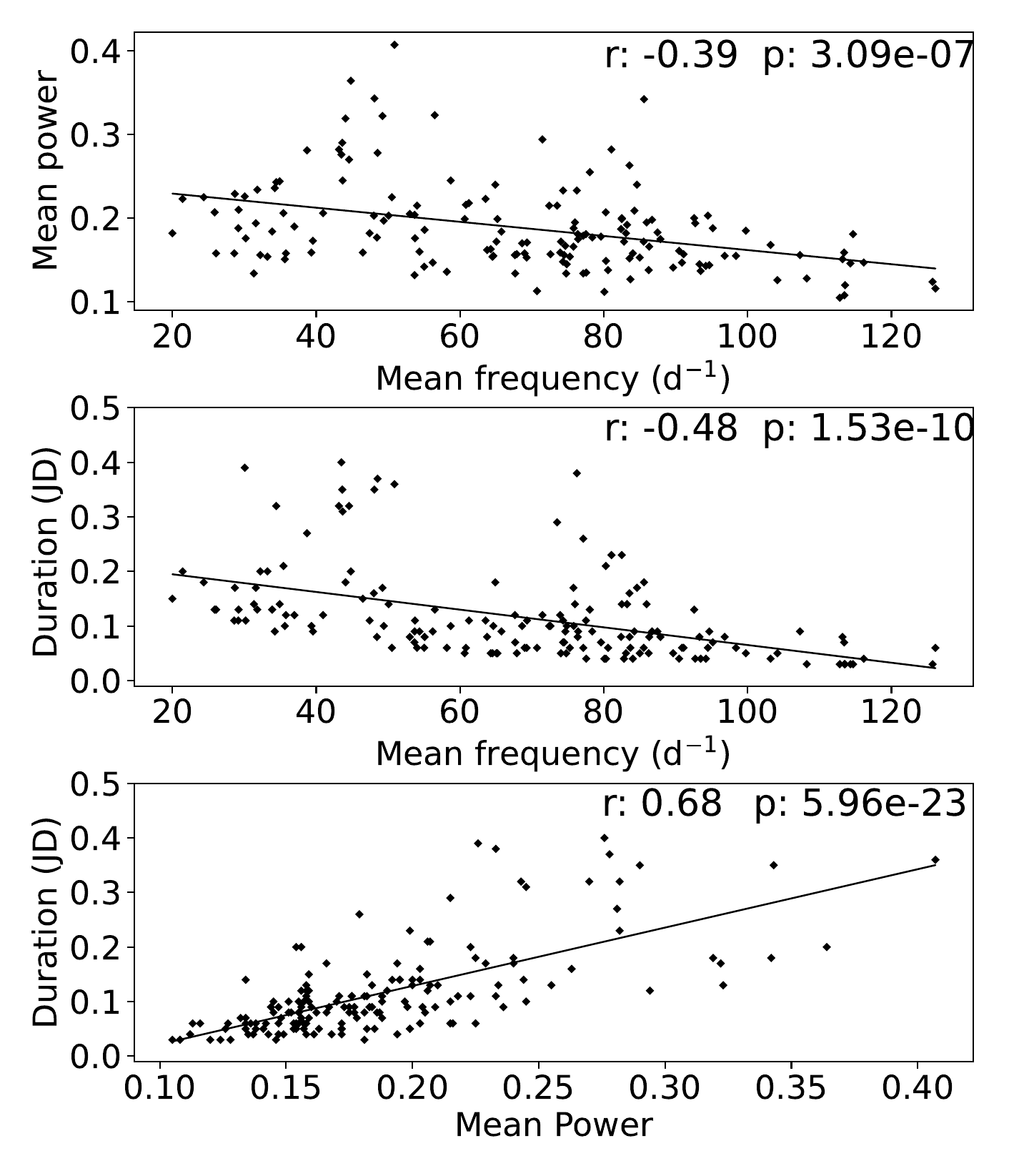}
    \caption{Correlation plots of the filtered data for $P \ge 0.10$. The panels compare, respectively: mean power and mean frequency (top), duration time of each group and their mean frequency (middle), and duration and mean power (bottom). The Pearson correlation coefficient (r) and the p-value for the null hypothesis (p) are included.
    }
    \label{fig:figure9}
\end{figure}

We find no correlation between the phase of the negative superhump and other quantities of QPOs, such as mean frequency (d$^{-1}$), group duration (JD), and number of cycles. \cite{2014AcA....64..167S} also shares this conclusion in his paper. We suspect that the physical cause of the QPOs is not related to the phase in any way.

Concerning the permanence of the oscillations, our research indicates that frequency groups usually persist for less than 7.5 hours. \cite{Y.Kim} 
noted significant variations in the period and semi-amplitude of QPOs between nights and within single observations, which is congruent with our time frames.
When considering completed cycles,
\cite{Kraicheva_1999}  
found that the oscillations remained coherent for about 3–8 cycles, while we find that most oscillations are coherent for between 3 and 18 cycles (see Fig. \ref{fig:figure6}). While this value is highly dependent on our requisites to form the groups, it also supports the idea that QPOs are brief and their frequencies dynamic. 

Regarding the variability of the oscillations, our frequency groups have relatively large frequency shifts, with a mean frequency amplitude of 9.8 d$^{-1}$. It could be possible for this value to be slightly larger if our requirements for forming frequency groups were less strict. Our results are comparable to those of \cite{Andronov_1999}, who find numerous frequency candidates with relatively large frequency shifts of up to 10 d$^{-1}$. 

\begin{figure}
        \includegraphics[width=\columnwidth]{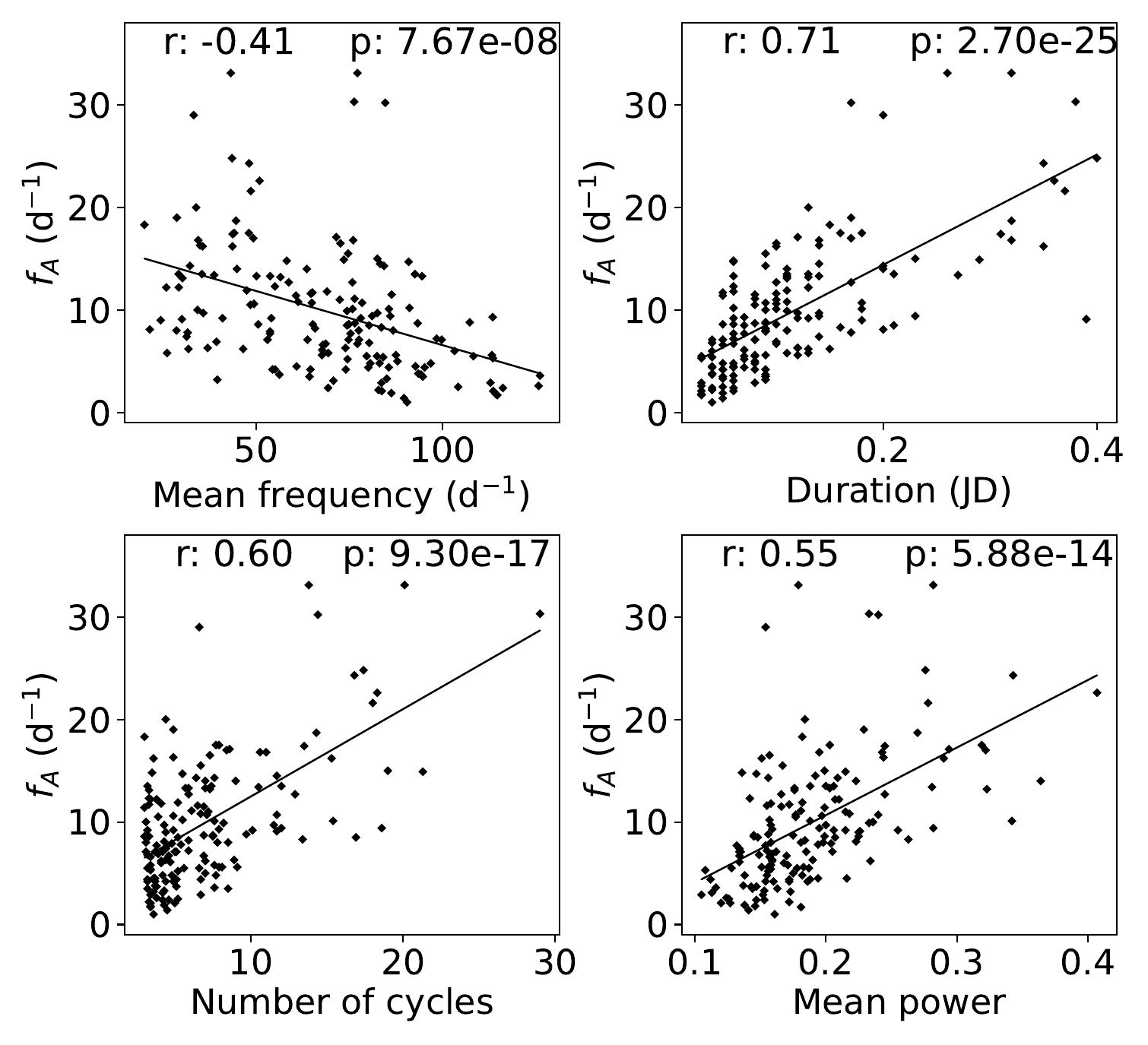}
    \caption{Correlation plot for $P \ge 0.10$, between the frequency amplitude ($f_{A}$) and four other quantities: mean frequency (d$^{-1}$) (top-left), duration (JD) (top-right), number of cycles (bottom-left), and mean power (bottom-right). The Pearson correlation coefficient (r) and the p-value for the null hypothesis (p) are included.
}
    \label{fig:figure10}
\end{figure}

\subsection{Flickering and its effects}
\label{sec:flick}
Besides QPOs, TT Ari presents `flickering', which is defined by \cite{1997ASPC..121..331W} 
as rapid, seemingly random fluctuations in light curves,  and has been seen in many astronomical sources. This phenomenon can be observed in any accretion-powered source, not just CVs. The fluctuations show similarities to noise, and other authors have studied them as if they were such. However, flickering is intrinsic to the source itself, unlike normal observational noise. Flickering amplitude and period depend on the individual system \citep{Bruch_2021}. 
On TT Ari, these fluctuations occur with an interval of minutes, and their amplitude is of the same order of magnitude as QPOs. This makes separating flickering from QPOs a rather difficult task. Other authors have expressed their doubts on the subject:
\cite{Warner_2004} 
note some of the flickering and flaring documented in the CV literature shows a resemblance to QPO, and \cite{Bruch_2014} 
asserts that an incidental superposition of unrelated flickering events occurring on similar timescales and at similar magnitude could mimic QPOs. Because of the difficulty in separating these two phenomena, we instead focused on measuring its effects on our groups, which we accomplished using randomised light curves, as described in Sect. \ref{sec:random_data}, with the results presented in Sect. \ref{sec:random_data_findings}. 

Our method can create frequency groups from randomised light curves. These groups can have a similar distribution to those found in the original data (see \ref{fig:figure11}). This highlights the possible effect of stochastic phenomena on QPOs. Despite this, the significant difference in the quantity and quality of groups between random and original data implies that only a fraction of the found groups could possibly be explained by flickering; the remaining groups must be evidence of QPOs. From Table \ref{table3_new}, we can quantify this fraction: random processes could account for up to 43\% of all frequency groups found and for up to 29\% of the individual frequencies. For a stricter filter of $P \ge 0.15$, these numbers decrease to 17\% and 11\%, respectively. 
To the best of our knowledge, our method has not been used before in this field of research.

\section{Conclusions}

Based on the results of this study, we have identified evidence suggesting the existence of QPOs. Our main findings can be summarised as follows:
   \begin{enumerate}
      \item QPOs are prevalent in TT Ari, and have non-random frequencies. The light curve of TT Ari presents one or more QPOs for around 78.3\% of the length of the observations studied here, most of them in the range of 25 to 90 d$^{-1}$ (16 to 53 minutes), with two peaks at 43.5 and 77.5 d$^{-1}$ (18.5 and 33 minutes). As the two peaks in frequency are not a multiple of one another and have different shapes, we do not believe that they are harmonics. There are very few groups with frequencies of over 100 d$^{-1}$, and they are unlikely to represent real QPOs. 
      \item QPOs shift rapidly. Most frequency groups persist for less than 7.5 hours and they can complete $\sim$7.0 cycles on average before shifting frequencies.
      \item There are time intervals in the observation where two or more groups appear simultaneously. For around 37\% of the observation time, we found two
or more QPOs existing simultaneously. This suggests that QPOs can exist at two different frequencies at the same time.
      \item No correlation was found between the superhump phase and other quantities such as mean frequency, frequency amplitude, or number of completed cycles of QPO.
      \item The difference in quantity and quality of the groups between randomised light curves and the original implies that most QPOs are not caused by random fluctuations, such as flickering. Stochastic processes could account for up to 43\% of the groups found with a less strict filter, but this decreases to 17\% when we use a stricter power threshold.

   \end{enumerate}
   
The same method used here for TT Ari can be applied to other targets, hopefully serving as an additional tool in the investigation of QPOs. However, this method requires attention to detail. When creating frequency groups, the filtering parameters must be set with care so that the groups can correctly identify QPOs. This is especially important for minimising false positives from random fluctuations.

\begin{acknowledgements}
      We would like to thank the team of the MOST space telescope, in particular A.F.J. Moffat, J.M. Matthews, R. Kuschnig, D.B. Guenther, J.F. Rowe, S.M. Rucinski, D. Sasselov and W.W. Weiss for assigning the observation time, executing the observations and sending us the reduced data set of the TT Ari campaign. We also thank the referee for very valuable comments on an earlier version of this article. NV  acknowledges financial support by Centro de Astrofísica de Valparaíso (CAV). 
\end{acknowledgements}

\bibliography{bibgraph}

@ARTICLE{Lomb,
       author = {{Lomb}, N.~R.},
        title = "{Least-Squares Frequency Analysis of Unequally Spaced Data}",
      journal = {\apss},
         year = 1976,
        month = feb,
       volume = {39},
       number = {2},
        pages = {447-462},
          doi = {10.1007/BF00648343},
       adsurl = {https://ui.adsabs.harvard.edu/abs/1976Ap&SS..39..447L}
}

@ARTICLE{Scargle,
       author = {{Scargle}, J.~D.},
        title = "{Studies in astronomical time series analysis. II. Statistical aspects of spectral analysis of unevenly spaced data.}",
      journal = {\apj},
         year = 1982,
        month = dec,
       volume = {263},
        pages = {835-853},
          doi = {10.1086/160554},
       adsurl = {https://ui.adsabs.harvard.edu/abs/1982ApJ...263..835S}
}

@article{Tremko_1996,
author = {Tremko, J. and Andronov, Ivan and Chinarova, L. and Kumsiashvili, M. and Luthardt, R. and Pajdosz, G. and Patkós, L. and Roessiger, S. and Zola, Stuart},
year = {1996},
month = {07},
pages = {121-134},
title = {Periodic and aperiodic variations in TT Arietis: Results from an international campaign},
volume = {312},
journal = {Astronomy and Astrophysics}
}

@article{Andronov_1999,
	doi = {10.1086/300665},
	url = {https://doi.org/10.1086/300665},
	year = 1999,
	month = {jan},
	publisher = {American Astronomical Society},
	volume = {117},
	number = {1},
	pages = {574--586},
	author = {I. L. Andronov and K. Arai and L. L. Chinarova and N. I. Dorokhov and T. N. Dorokhova and A. Dumitrescu and D. Nogami and S. V. Kolesnikov and A. Lepardo and P. A. Mason and K. Matsumoto and G. Oprescu and G. Pajdosz and R. Passuelo and L. Patkos and D. S. Senio and G. Sostero and V. F. Suleimanov and J. Tremko and G. V. Zhukov and S. Zola},
	title = {A Search for Periodic and Quasi-periodic Photometric Behavior in the Cataclysmic Variable {TT} Arietis},
	journal = {The Astronomical Journal}
}

@ARTICLE{Kraicheva_1999,
       author = {{Kraicheva}, Z. and {Stanishev}, V. and {Genkov}, V. and {Iliev}, L.},
        title = "{TT Arietis: 1985-1999 accretion disc behaviour}",
      journal = {\aap},
         year = 1999,
        month = nov,
       volume = {351},
        pages = {607-618},
       adsurl = {https://ui.adsabs.harvard.edu/abs/1999A&A...351..607K}
}

@article{Bruch_2019,
author = {Bruch, Albert},
year = {2019},
month = {10},
pages = {2961-2975},
title = {TT Arietis: 40 yr of photometry},
volume = {489},
journal = {Monthly Notices of the Royal Astronomical Society},
doi = {10.1093/mnras/stz2381}
}

@ARTICLE{2014AcA....64..167S,
       author = {{Smak}, J.},
        title = "{TT Ari and its Quasi-Periodic Oscillations}",
      journal = {\actaa},
         year = 2014,
        month = jun,
       volume = {64},
       number = {2},
        pages = {167-175},
archivePrefix = {arXiv},
       eprint = {1404.6349},
 primaryClass = {astro-ph.SR},
       adsurl = {https://ui.adsabs.harvard.edu/abs/2014AcA....64..167S}
}

@ARTICLE{1988AcA....38..315U,
       author = {{Udalski}, A.},
        title = "{Photometry of cataclysmic variables. II. TT Arietis.}",
      journal = {\actaa},
         year = 1988,
        month = jan,
       volume = {38},
        pages = {315-327},
       adsurl = {https://ui.adsabs.harvard.edu/abs/1988AcA....38..315U}
}

@ARTICLE{Hollander_1992,
       author = {{Hollander}, A. and {van Paradijs}, J.},
        title = "{Quasi-periodic oscillations in TT Arietis.}",
      journal = {\aap},
         year = 1992,
        month = nov,
       volume = {265},
        pages = {77-81},
       adsurl = {https://ui.adsabs.harvard.edu/abs/1992A&A...265...77H}
}

@ARTICLE{Semeniuk_1987,
       author = {{Semeniuk}, I. and {Schwarzenberg-Czerny}, A. and {Duerbeck}, H. and {Hoffmann}, M. and {Smak}, J. and {Stepien}, K. and {Tremko}, J.},
        title = "{Photometry of TT Arietis}",
      journal = {\apss},
         year = 1987,
        month = feb,
       volume = {130},
       number = {1-2},
        pages = {167-174},
          doi = {10.1007/BF00654990},
       adsurl = {https://ui.adsabs.harvard.edu/abs/1987Ap&SS.130..167S}
}

@article{Y.Kim,
	author = {{Kim, Y.} and {Andronov, I. L.} and {Cha, S. M.} and {Chinarova, L. L.} and {Yoon, J. N.}},
	title = {Nova-like cataclysmic variable TT Arietis* - QPO behaviour coming back from positive superhumps},
	DOI= "10.1051/0004-6361:200810005",
	url= "https://doi.org/10.1051/0004-6361:200810005",
	journal = {A\&A},
	year = 2009,
	volume = 496,
	number = 3,
	pages = "765-775",
}

@ARTICLE{William_1966,
       author = {{Williams}, John O.},
        title = "{A Peculiar Beat Phenomenon in the Light Variation of BD +14{\textdegree}341}",
      journal = {\pasp},
         year = 1966,
        month = aug,
       volume = {78},
       number = {464},
        pages = {279},
          doi = {10.1086/128349},
       adsurl = {https://ui.adsabs.harvard.edu/abs/1966PASP...78..279W}
}

@article{Vogt_2013,
	doi = {10.1002/asna.201311949},
  
	url = {https://doi.org/10.1002%2Fasna.201311949},
  
	year = 2013,
	month = {dec},
  
	publisher = {Wiley},
  
	volume = {334},
  
	number = {10},
  
	pages = {1101--1106},
  
	author = {N. Vogt and A.-N. Chen{\'{e}
} and A.F.J. Moffat and J.M. Matthews and R. Kuschnig and D.B. Guenther and J.F. Rowe and S.M. Rucinski and D. Sasselov and W.W. Weiss},
  
	title = {A photometric study of the nova-like variable {TT} Arietis with the {MOST} satellite},
  
	journal = {Astronomische Nachrichten}
}

@ARTICLE{2011ApJ...735...34C,
       author = {{Chen{\'e}}, A. -N. and {Moffat}, A.~F.~J. and {Cameron}, C. and {Fahed}, R. and {Gamen}, R.~C. and {Lef{\`e}vre}, L. and {Rowe}, J.~F. and {St-louis}, N. and {Muntean}, V. and {De La Chevroti{\`e}re}, A. and {Guenther}, D.~B. and {Kuschnig}, R. and {Matthews}, J.~M. and {Rucinski}, S.~M. and {Sasselov}, D. and {Weiss}, W.~W.},
        title = "{WR 110: A Single Wolf-Rayet Star with Corotating Interaction Regions in its Wind?}",
      journal = {\apj},
         year = 2011,
        month = jul,
       volume = {735},
       number = {1},
          eid = {34},
        pages = {34},
          doi = {10.1088/0004-637X/735/1/34},
archivePrefix = {arXiv},
       eprint = {1105.0919},
 primaryClass = {astro-ph.SR},
       adsurl = {https://ui.adsabs.harvard.edu/abs/2011ApJ...735...34C}
}

@ARTICLE{1999JRASC..93..183M,
       author = {{Matthews}, J. and {Kuschnig}, R. and {Walker}, G. and {Johnson}, R. and {Skaret}, K. and {Shkolnik}, E. and {Lanting}, T. and {Morgan}, J.~P. and {Pazder}, J. and {Sinclair}, P. and {Harron}, J. and {Sturgeon}, D.},
        title = "{The MOST space mission: a 15-cm telescope in the 8-m-class era.}",
      journal = {\jrasc},
         year = 1999,
        month = aug,
       volume = {93},
       number = {4},
        pages = {183-184},
       adsurl = {https://ui.adsabs.harvard.edu/abs/1999JRASC..93..183M}
}

@ARTICLE{2003PASP..115.1023W,
       author = {{Walker}, Gordon and {Matthews}, Jaymie and {Kuschnig}, Rainer and {Johnson}, Ron and {Rucinski}, Slavek and {Pazder}, John and {Burley}, Gregory and {Walker}, Andrew and {Skaret}, Kristina and {Zee}, Robert and {Grocott}, Simon and {Carroll}, Kieran and {Sinclair}, Peter and {Sturgeon}, Don and {Harron}, John},
        title = "{The MOST Asteroseismology Mission: Ultraprecise Photometry from Space}",
      journal = {\pasp},
         year = 2003,
        month = sep,
       volume = {115},
       number = {811},
        pages = {1023-1035},
          doi = {10.1086/377358},
       adsurl = {https://ui.adsabs.harvard.edu/abs/2003PASP..115.1023W}
}

@INPROCEEDINGS{1999dicb.conf...61P,
       author = {{Patterson}, J.},
        title = "{Permanent Superhumps in Cataclysmic Variables}",
    booktitle = {Disk Instabilities in Close Binary Systems},
         year = 1999,
       editor = {{Mineshige}, S. and {Wheeler}, J.~C.},
        month = jan,
        pages = {61},
       adsurl = {https://ui.adsabs.harvard.edu/abs/1999dicb.conf...61P}
}

@INPROCEEDINGS{1997ASPC..121..331W,
       author = {{Welsh}, W.~F. and {Wood}, J.~H. and {Horne}, K.},
        title = "{Recent Observations of Flickering in CVs and AGN}",
    booktitle = {IAU Colloq. 163: Accretion Phenomena and Related Outflows},
         year = 1997,
       editor = {{Wickramasinghe}, D.~T. and {Bicknell}, G.~V. and {Ferrario}, L.},
       series = {Astronomical Society of the Pacific Conference Series},
       volume = {121},
        month = jan,
        pages = {331},
       adsurl = {https://ui.adsabs.harvard.edu/abs/1997ASPC..121..331W}
}

@article{Warner_2004,
	doi = {10.1086/381742},
	url = {https://doi.org/10.1086/381742},
	year = 2004,
	month = {feb},
	publisher = {{IOP} Publishing},
	volume = {116},
	number = {816},
	pages = {115--132},
	author = {Brian Warner},
	title = {Rapid Oscillations in Cataclysmic Variables},
	journal = {Publications of the Astronomical Society of the Pacific}
}

@article{Bruch_2021,
	doi = {10.1093/mnras/stab516},
  
	url = {https://doi.org/10.1093%2Fmnras%2Fstab516},
  
	year = 2021,
	month = {feb},
  
	publisher = {Oxford University Press ({OUP})},
  
	volume = {503},
  
	number = {1},
  
	pages = {953--971},
  
	author = {Albert Bruch},
  
	title = {A comparative study of the strength of flickering in cataclysmic variables},
  
	journal = {Monthly Notices of the Royal Astronomical Society}
}

@ARTICLE{Bruch_2014,
       author = {{Bruch}, Albert},
        title = "{Long-term photometry of the eclipsing dwarf nova V893 Scorpii. Orbital period, oscillations, and a possible giant planet}",
      journal = {\aap},
         year = 2014,
        month = jun,
       volume = {566},
          eid = {A101},
        pages = {A101},
          doi = {10.1051/0004-6361/201423576},
archivePrefix = {arXiv},
       eprint = {1404.2902},
 primaryClass = {astro-ph.SR},
       adsurl = {https://ui.adsabs.harvard.edu/abs/2014A&A...566A.101B}
}

@MISC{aavso,
  year=2018,
  title={{AAVSO} Photometric All Sky Survey ({APASS}) DR10},
  author={Henden, Arne A. and Levine, Stephen and Terrell, Dirk and Welch, Douglas L. and Munari, Ulisse and Kloppenborg, Brian K.},
  url={https://dc.zah.uni-heidelberg.de/apass/q/cone/info},
  howpublished={{VO} resource provided by the {GAVO} Data Center}
}

@inproceedings{VanderPlas_2012,
   title={Introduction to astroML: Machine learning for astrophysics},
   volume={195},
   url={http://dx.doi.org/10.1109/CIDU.2012.6382200},
   DOI={10.1109/cidu.2012.6382200},
   booktitle={2012 Conference on Intelligent Data Understanding},
   publisher={IEEE},
   author={VanderPlas, Jacob and Connolly, Andrew J. and Ivezic, Zeljko and Gray, Alex},
   year={2012},
   month=oct, pages={47–54} }

@ARTICLE{VanderPlas_2015,
       author = {{VanderPlas}, Jacob T. and {Ivezi{\'c}}, {\v{Z}}eljko},
        title = "{Periodograms for Multiband Astronomical Time Series}",
      journal = {\apj},
         year = 2015,
        month = oct,
       volume = {812},
       number = {1},
          eid = {18},
        pages = {18},
          doi = {10.1088/0004-637X/812/1/18},
archivePrefix = {arXiv},
       eprint = {1502.01344},
 primaryClass = {astro-ph.IM},
       adsurl = {https://ui.adsabs.harvard.edu/abs/2015ApJ...812...18V}
}

\begin{appendix}
\onecolumn
\section{Figures}
\begin{figure}[hbp]
\centering
\begin{subfigure}{}
    \includegraphics[width=0.45\columnwidth]{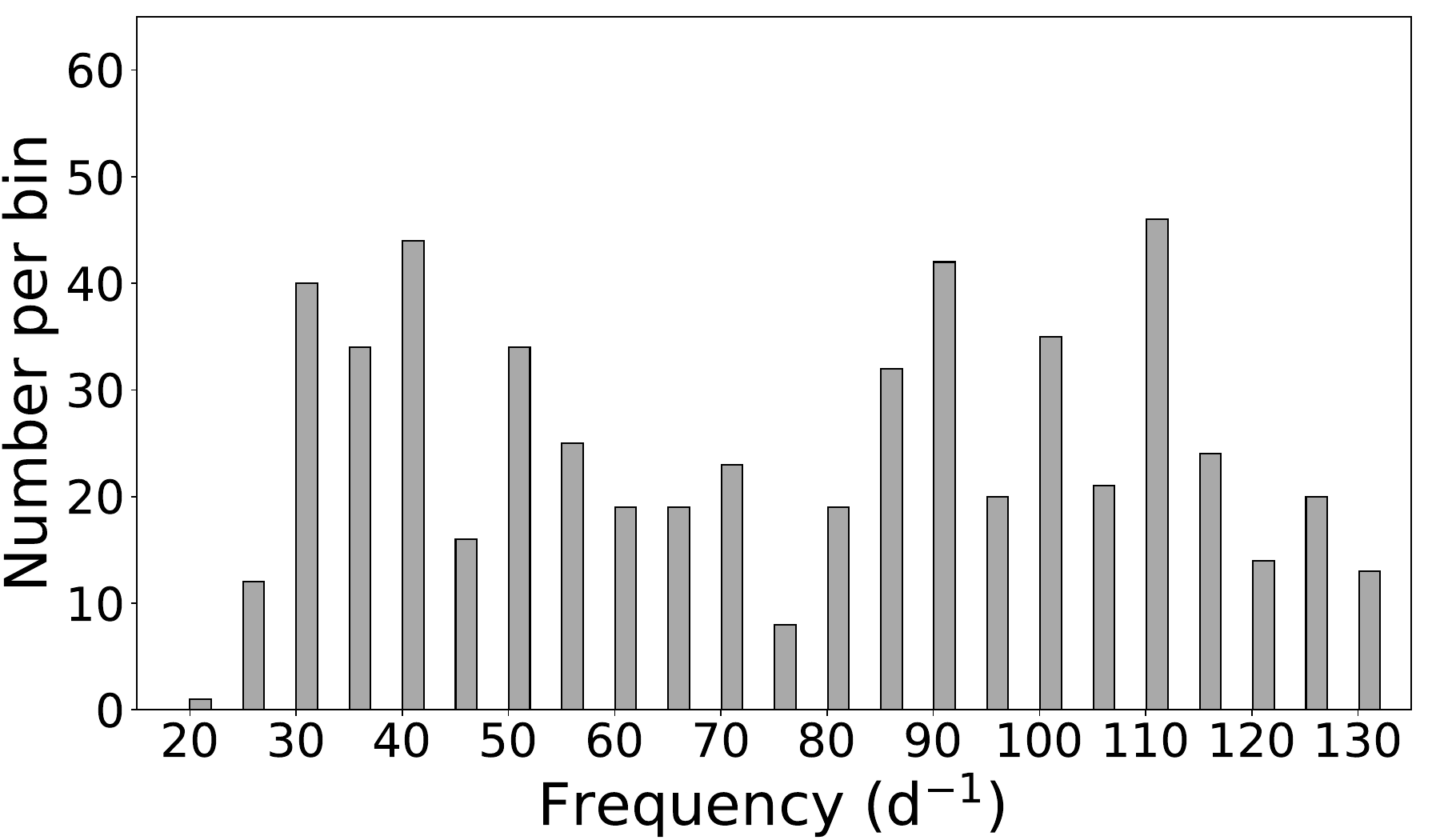}
    \end{subfigure}
    \vspace{-0.3\baselineskip}
\begin{subfigure}{}
        \includegraphics[width=0.45\columnwidth]{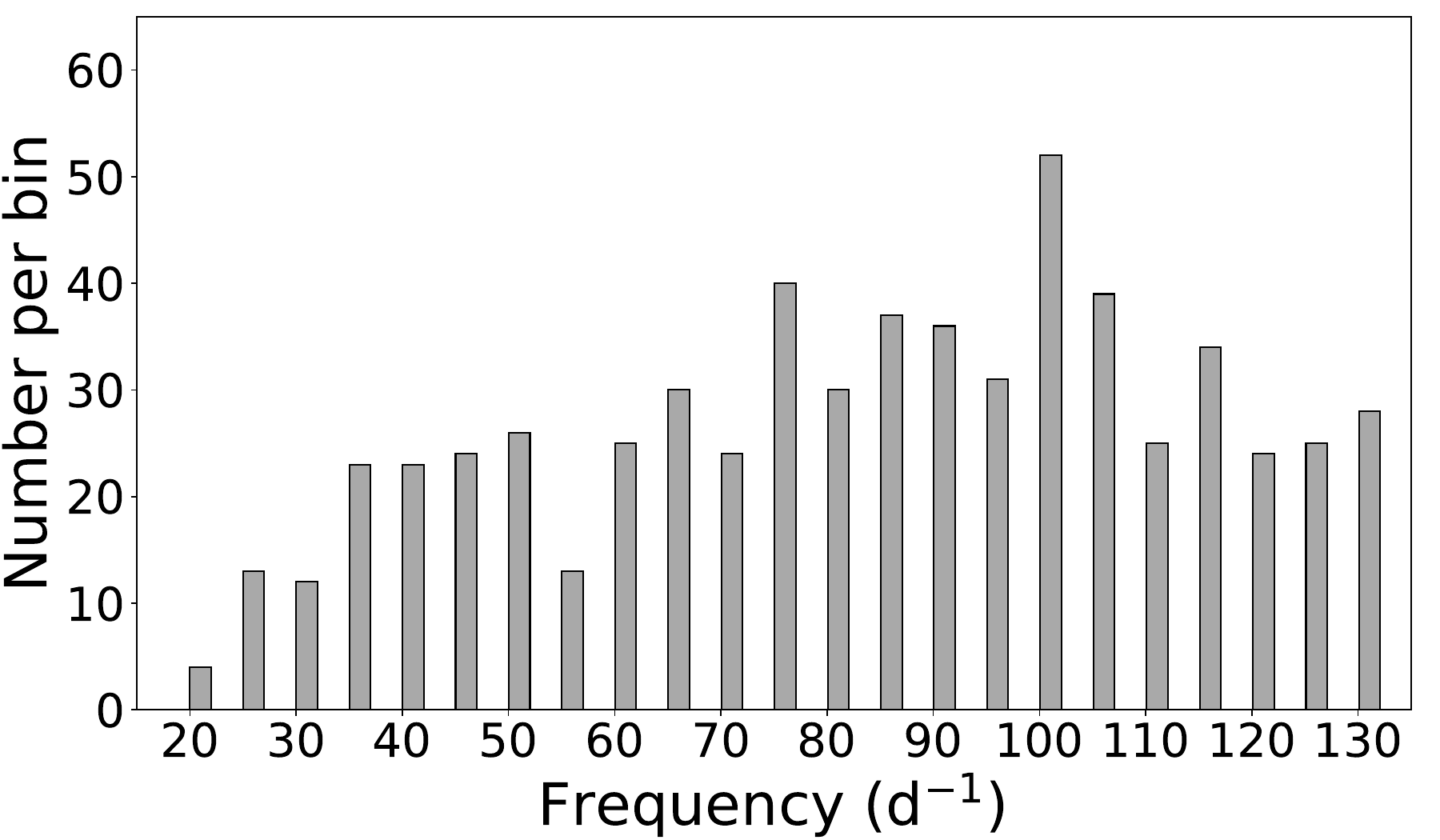}
    \end{subfigure}
    \vspace{-0.3\baselineskip}    
\begin{subfigure}{}
        \includegraphics[width=0.45\columnwidth]{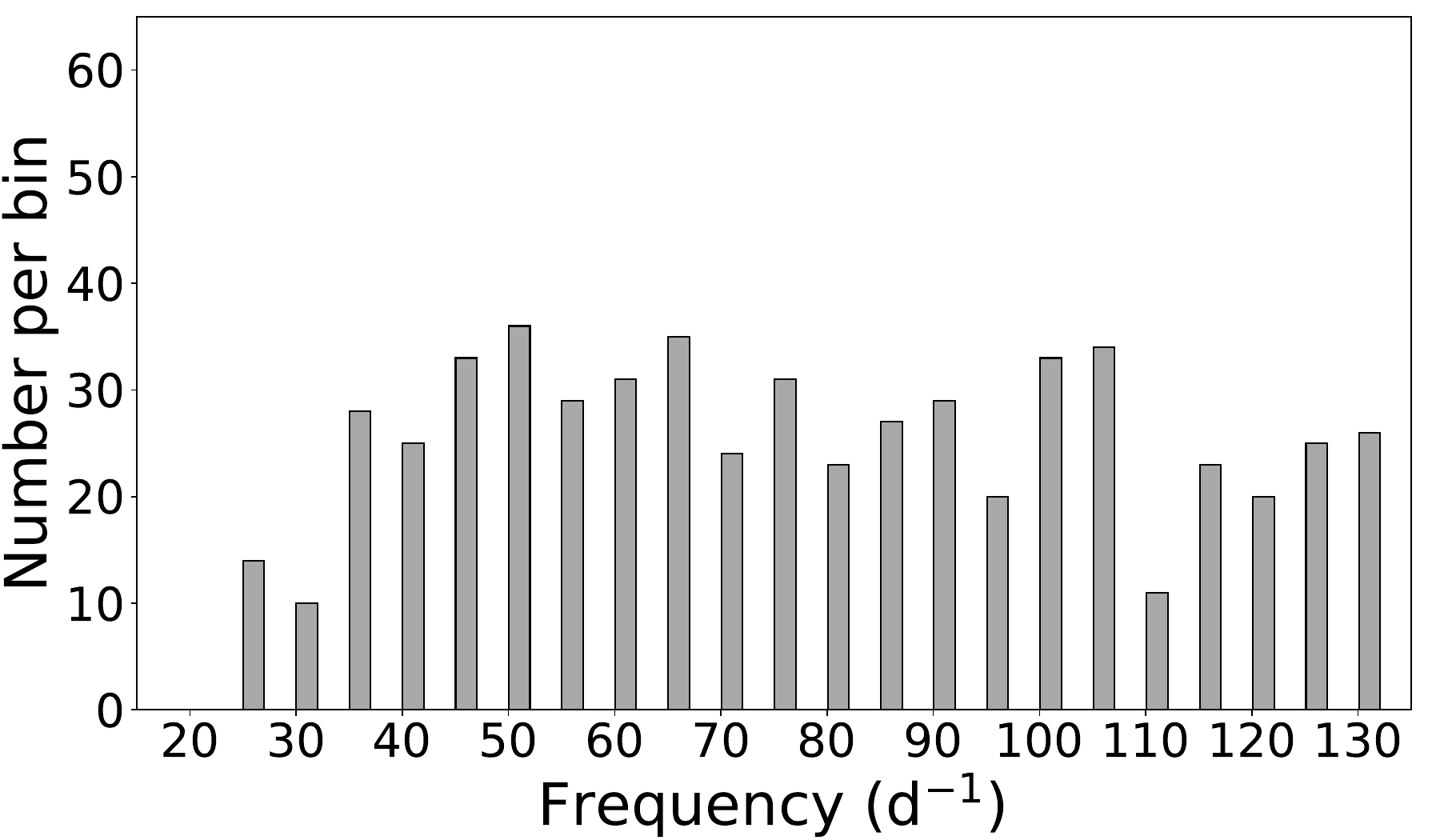}
    \end{subfigure}
    \vspace{-0.2\baselineskip}
\begin{subfigure}{}
        \includegraphics[width=0.45\columnwidth]{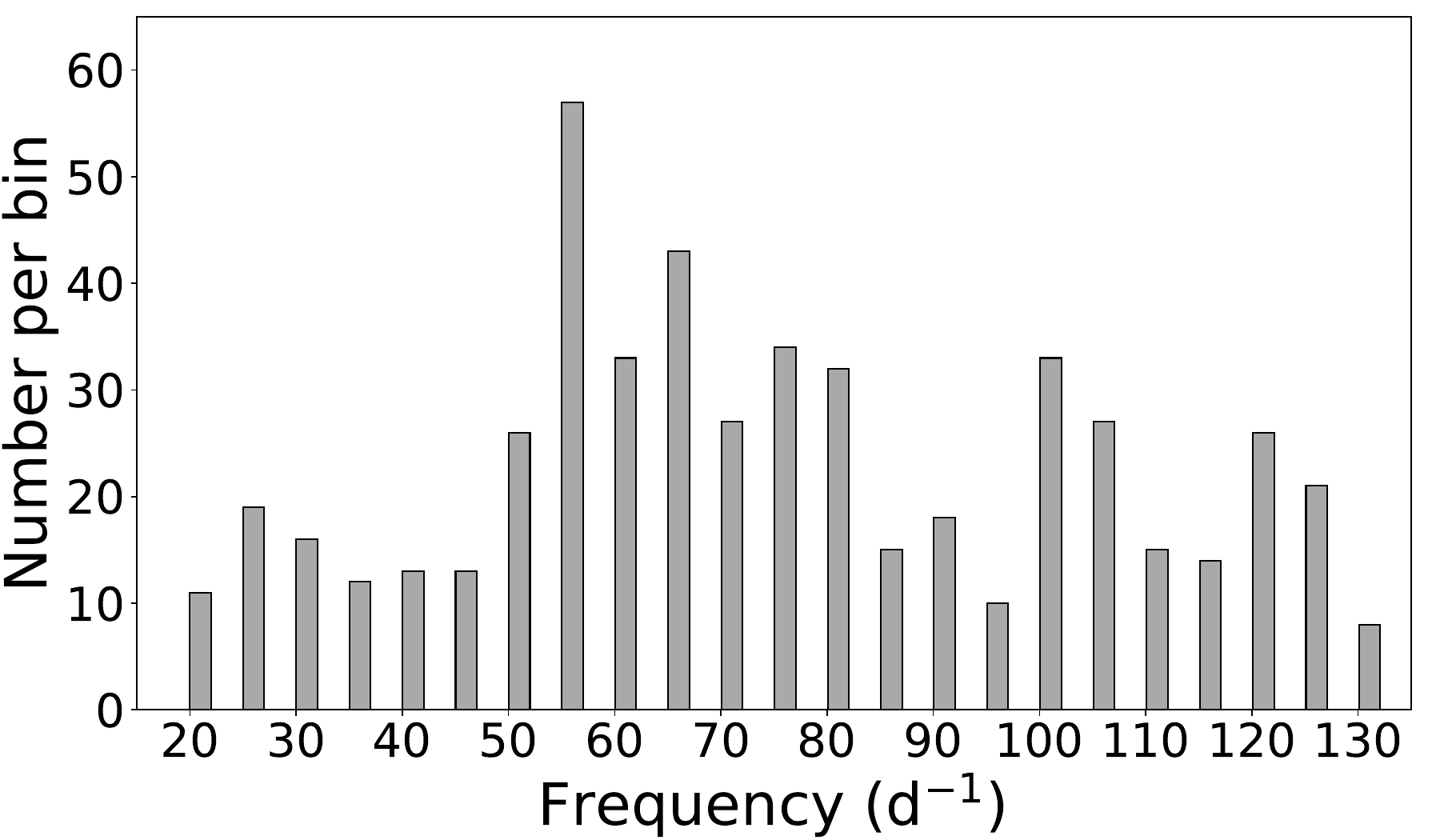}
    \end{subfigure}
    \vspace{-0.2\baselineskip}
\begin{subfigure}{}
        \includegraphics[width=0.45\columnwidth]{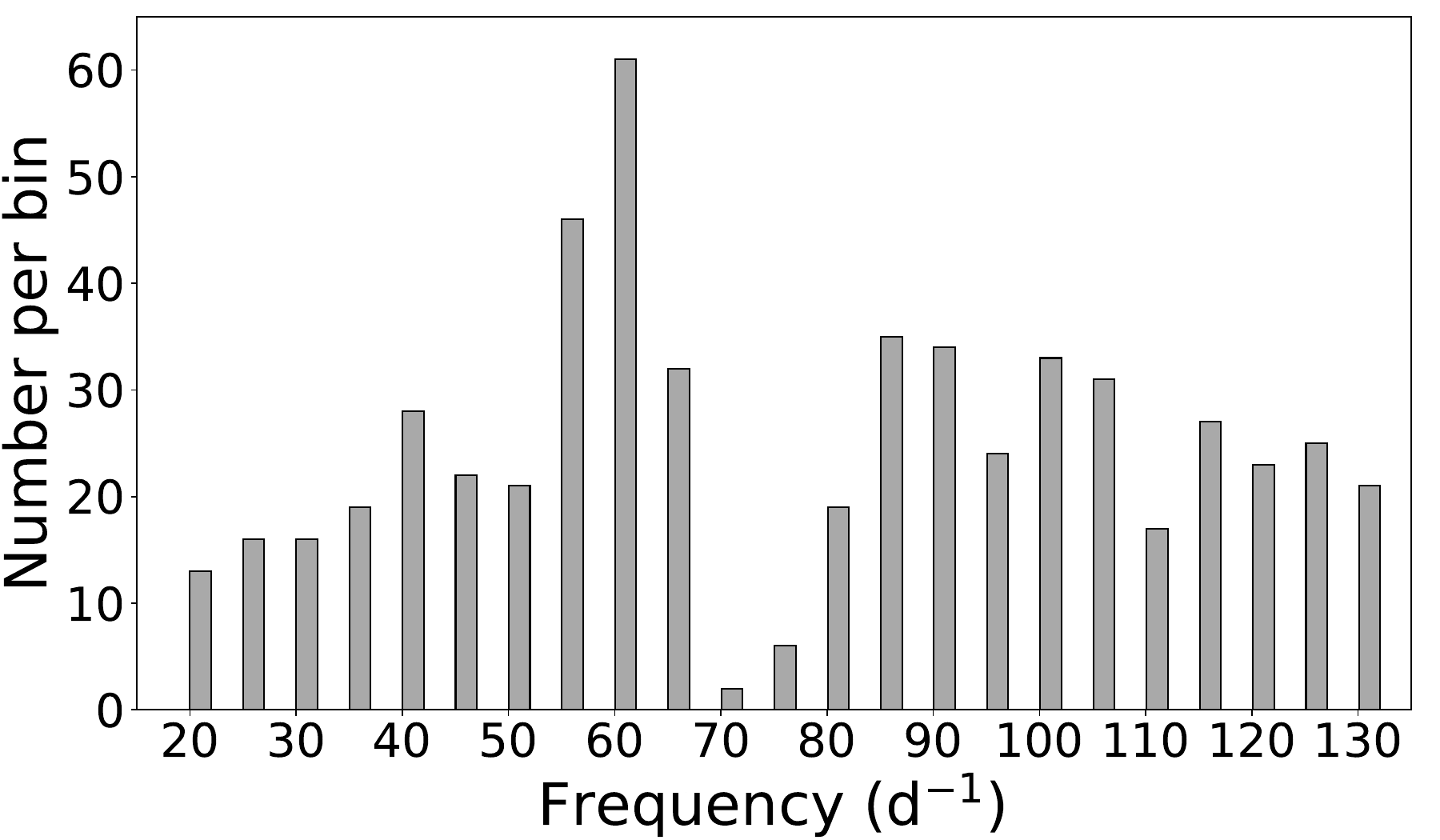}
    \end{subfigure}
    \vspace{-0.2\baselineskip}
\begin{subfigure}{}
        \includegraphics[width=0.45\columnwidth]{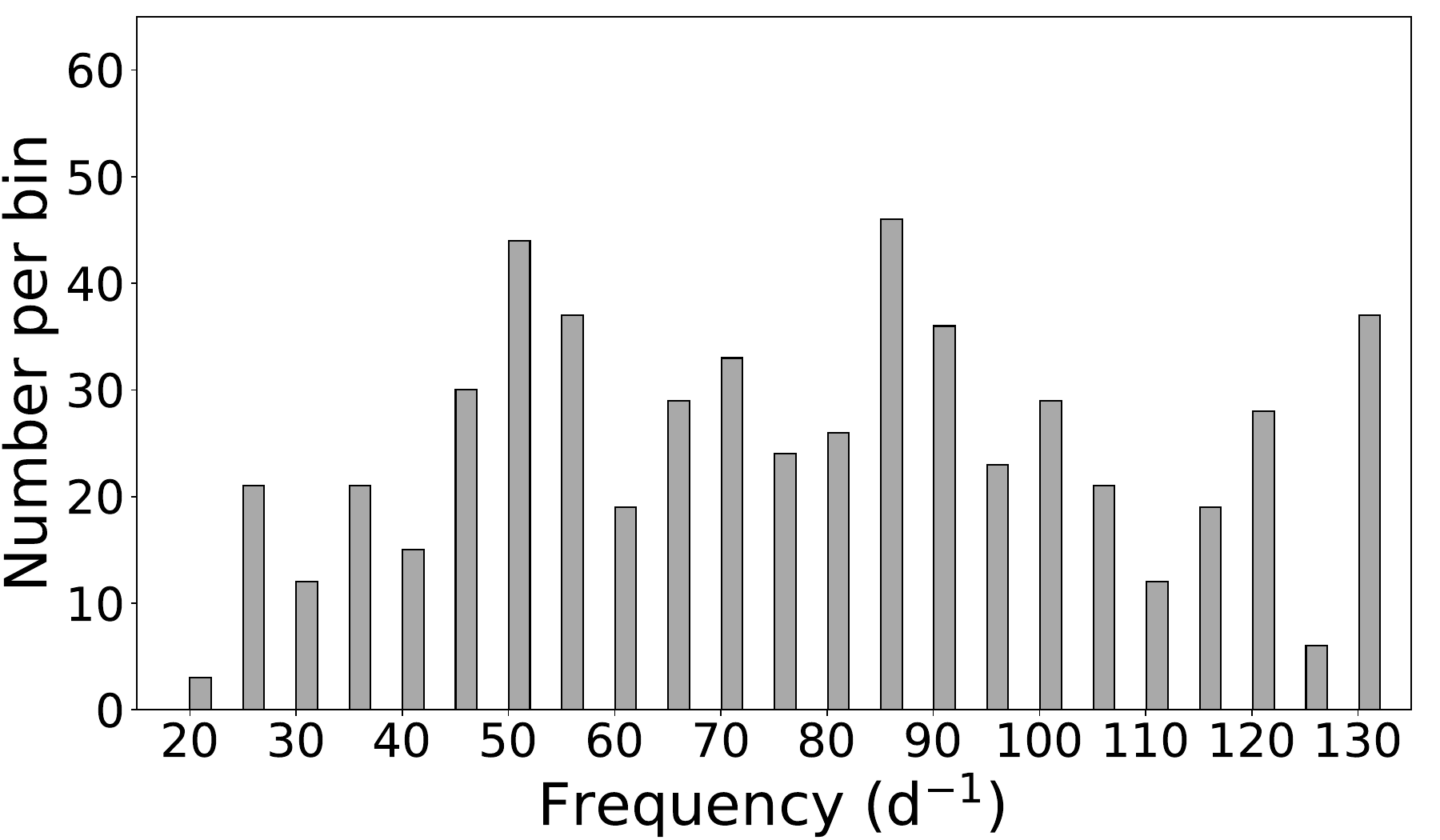}
    \end{subfigure}
    \vspace{-0.2\baselineskip}
\begin{subfigure}{}
        \includegraphics[width=0.45\columnwidth]{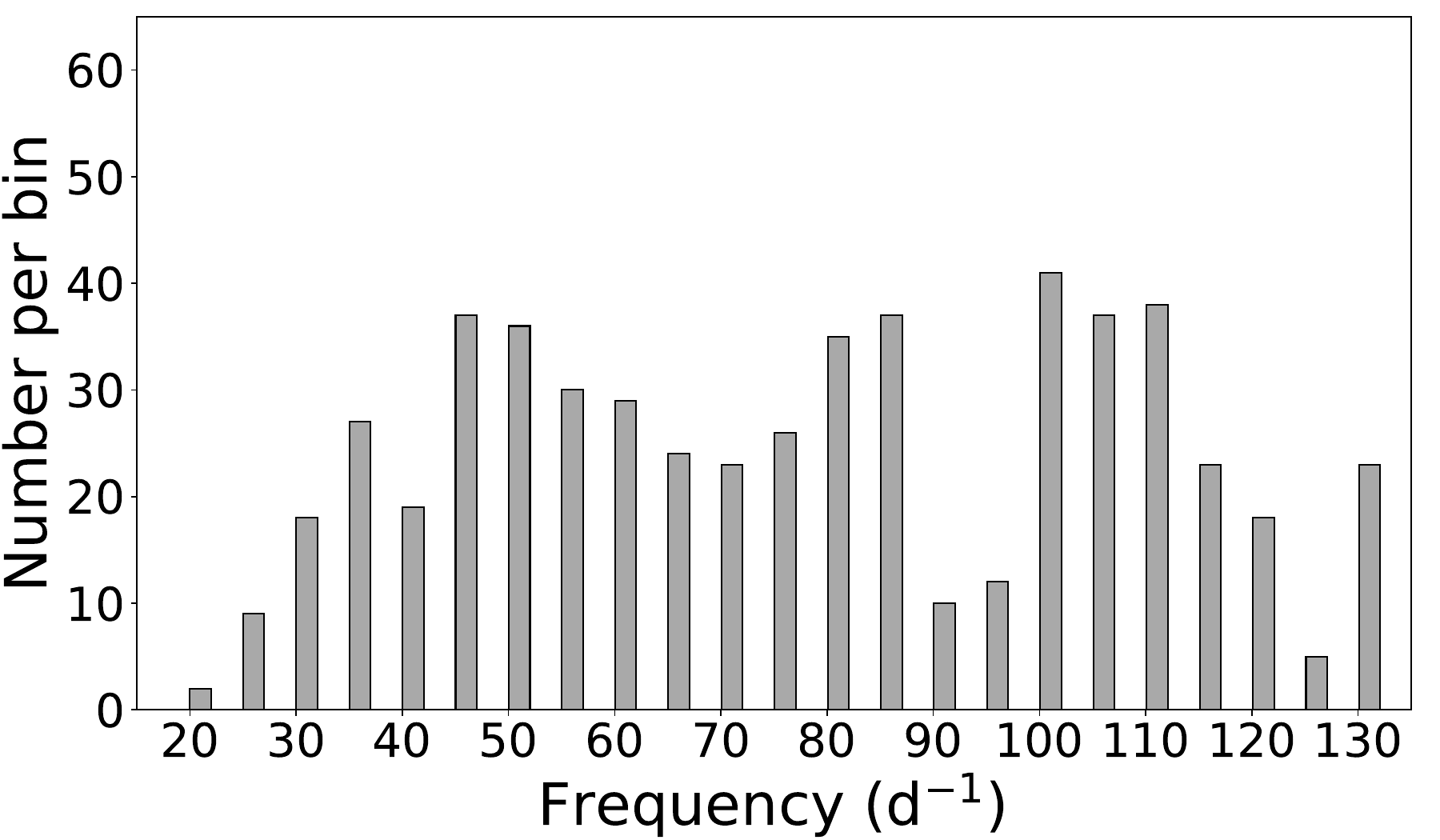}
    \end{subfigure}
\begin{subfigure}{}
        \includegraphics[width=0.45\columnwidth]{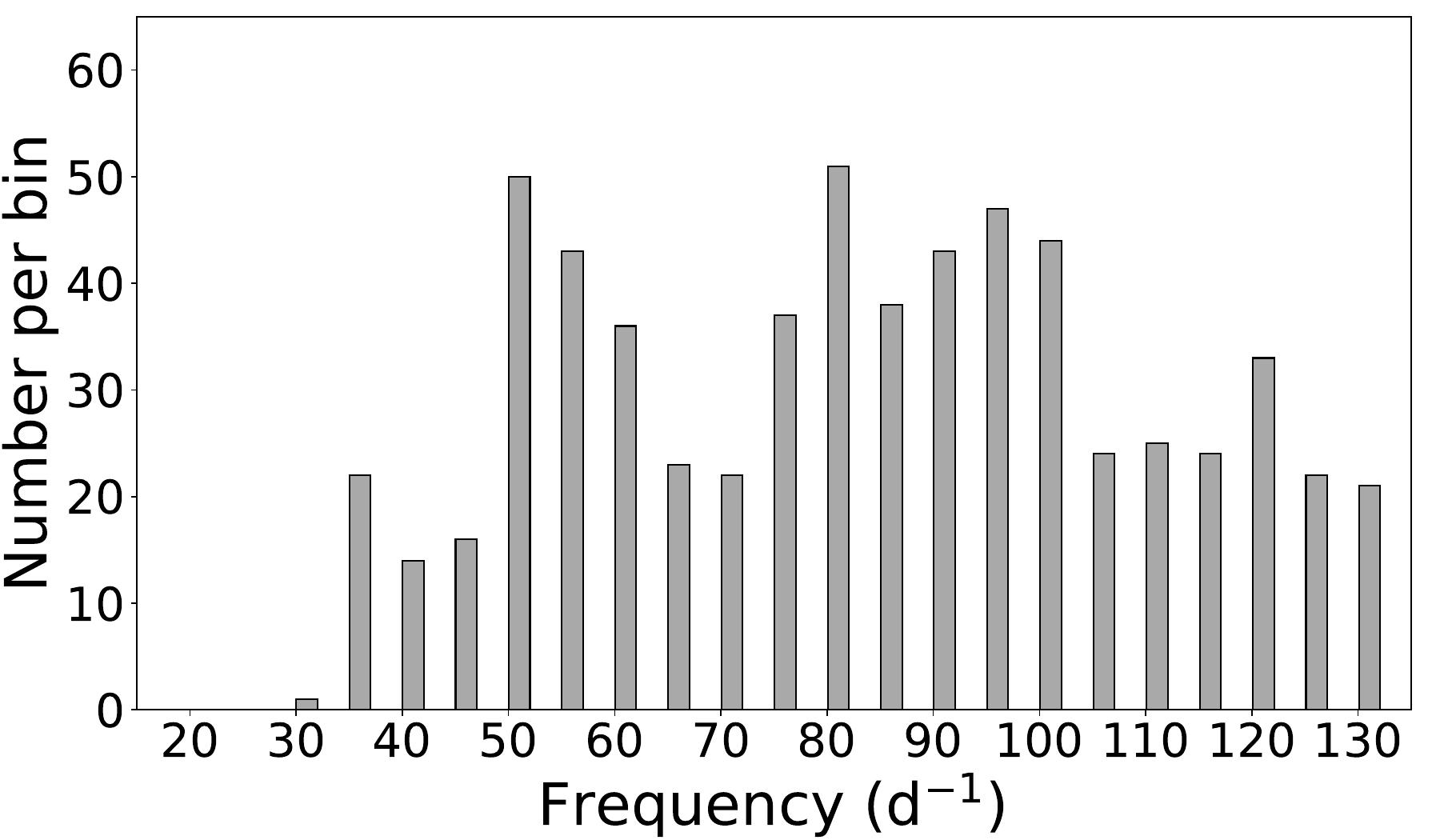}
    \end{subfigure}
\caption{
Frequency histograms for groups made from randomised light curves, for $P \ge 0.10$ in the power spectrum. Each panel corresponds to a different iteration of the random shuffling process.
}
\label{fig:figure11}
\end{figure}

\clearpage
\newpage
\section{Tables}


\begin{longtable}[c]{p{1.2cm}rccrrrrl} 
\caption{Detailed information of every frequency group found, made from frequencies with  high power (P) in the power spectrum, $P \ge 0.10$ in this case. From the columns: Starting date shows the starting time of each group, the numbers displayed are JD +2456000. $\overline{P}$ is the mean power (of each group) in the power spectrum. The phase was calculated for the superhump maximum. Quantities such as $\overline{f}$, duration, $f_{\Delta}$, $f_{A}$ and $N_{c}$ are described in Table \ref{tab:table2}.}
\label{tab:table3}\\
\hline
\begin{tabular}[l]{@{}l@{}}Starting \\ date (JD)\\  \end{tabular} & \begin{tabular}[l]{@{}l@{}} $\overline{f}$ (d$^{-1}$)\end{tabular} & \begin{tabular}[l]{@{}l@{}}Duration \\(JD)\end{tabular} & \begin{tabular}[l]{@{}l@{}}$\overline{P}$\end{tabular} & \begin{tabular}[l]{@{}l@{}}$f_{\Delta}$  (d$^{-1}$)\end{tabular} & \begin{tabular}[l]{@{}l@{}}$f_{A}$ (d$^{-1}$) \end{tabular} & \begin{tabular}[l]{@{}l@{}}$N_{c}$\end{tabular} & Phase \\ 
\hline
\endfirsthead
\multicolumn{8}{c}%
{{\bfseries Table \thetable\ continued from previous page}} \\
\hline
\begin{tabular}[l]{@{}l@{}}Starting \\ date   (JD)\end{tabular} & \begin{tabular}[l]{@{}l@{}}$\overline{f}$ (d$^{-1}$)\end{tabular} & \begin{tabular}[r]{@{}l@{}}Duration\\(JD) \end{tabular} & \begin{tabular}[r]{@{}l@{}}$\overline{P}$\end{tabular} & \begin{tabular}[l]{@{}l@{}}$f_{\Delta}$ (d$^{-1}$)\end{tabular} & \begin{tabular}[l]{@{}l@{}}$f_{A}$ (d$^{-1}$) \end{tabular} & \begin{tabular}[l]{@{}l@{}}$N_{c}$\end{tabular} & Phase \\ 
\hline
\endhead
\hline
\endfoot
\endlastfoot
206.06 & 100.79 & 0.04 & 0.206 & 2.3 & 2.5 & 4.0 & 0.1693 \\
206.06 & 80.30 & 0.04 & 0.149 & 6.5 & 6.8 & 3.2 & 0.1693 \\
206.08 & 43.63 & 0.35 & 0.290 & -6.7 & 16.2 & 15.3 & 0.4861 \\
206.20 & 86.31 & 0.08 & 0.166 & 7.6 & 11.5 & 6.9 & 0.3732 \\
206.28 & 63.56 & 0.11 & 0.223 & 10.9 & 14.0 & 7.0 & 0.0881 \\
206.44 & 67.64 & 0.12 & 0.156 & -5.2 & 5.6 & 8.1 & 0.3296 \\
206.56 & 85.95 & 0.14 & 0.195 & 2.9 & 9.4 & 12.0 & 0.3078 \\
206.62 & 65.07 & 0.05 & 0.172 & -11.6 & 11.7 & 3.3 & 0.4207 \\
206.62 & 50.89 & 0.36 & 0.407 & 7.9 & 22.6 & 18.3 & 0.5870 \\
206.74 & 98.38 & 0.06 & 0.155 & 6.6 & 7.2 & 5.9 & 0.3613 \\
206.96 & 90.87 & 0.06 & 0.147 & -14.2 & 14.7 & 5.5 & 0.0167 \\
207.01 & 47.43 & 0.11 & 0.182 & 9.2 & 11.9 & 5.2 & 0.5811 \\
207.10 & 95.16 & 0.07 & 0.188 & -2.8 & 4.4 & 6.7 & 0.1078 \\
207.10 & 114.66 & 0.03 & 0.181 & -0.9 & 1.7 & 3.4 & 0.9573 \\
207.14 & 74.03 & 0.05 & 0.172 & -1.1 & 4.2 & 3.7 & 0.3335 \\
207.14 & 44.57 & 0.32 & 0.270 & 4.5 & 18.7 & 14.3 & 0.3493 \\
207.15 & 116.15 & 0.04 & 0.147 & -2.4 & 2.4 & 4.6 & 0.3711 \\
207.22 & 103.19 & 0.04 & 0.168 & -5.1 & 6.0 & 4.1 & 0.8979 \\
207.23 & 60.65 & 0.05 & 0.199 & 11.4 & 11.4 & 3.0 & 0.0107 \\
207.27 & 113.42 & 0.07 & 0.159 & 9.3 & 9.3 & 7.9 & 0.3870 \\
207.40 & 80.08 & 0.04 & 0.112 & 4.2 & 4.4 & 3.2 & 0.2523 \\
207.53 & 49.22 & 0.17 & 0.322 & -14.9 & 17.0 & 8.4 & 0.7196 \\
207.58 & 72.40 & 0.10 & 0.215 & 11.0 & 11.0 & 7.2 & 0.8325 \\
207.66 & 94.19 & 0.04 & 0.143 & 3.3 & 3.7 & 3.8 & 0.2087 \\
207.74 & 68.99 & 0.06 & 0.158 & -11.8 & 11.8 & 4.1 & 0.8859 \\
207.74 & 44.82 & 0.20 & 0.364 & 9.0 & 14.0 & 9.0 & 0.4126 \\
207.81 & 74.84 & 0.10 & 0.145 & 6.8 & 8.6 & 7.5 & 0.5631 \\
208.00 & 60.82 & 0.06 & 0.216 & 2.1 & 4.5 & 3.6 & 0.8423 \\
208.01 & 96.81 & 0.08 & 0.155 & -3.1 & 4.8 & 7.7 & 0.9928 \\
208.06 & 54.03 & 0.06 & 0.215 & 5.9 & 9.2 & 3.2 & 0.2938 \\
208.08 & 44.08 & 0.18 & 0.319 & 12.4 & 17.5 & 7.9 & 0.8958 \\
208.21 & 65.77 & 0.09 & 0.184 & -5.3 & 8.2 & 5.9 & 0.5354 \\
208.31 & 43.17 & 0.32 & 0.282 & 17.2 & 33.1 & 13.8 & 0.1532 \\
208.41 & 85.59 & 0.18 & 0.342 & -7.4 & 10.1 & 15.4 & 0.3789 \\
208.62 & 91.09 & 0.06 & 0.157 & 8.0 & 10.2 & 5.5 & 0.5076 \\
208.67 & 48.54 & 0.37 & 0.278 & -11.7 & 21.6 & 18.0 & 0.0501 \\
208.67 & 76.37 & 0.08 & 0.181 & -5.1 & 11.1 & 6.1 & 0.9591 \\
208.70 & 84.61 & 0.17 & 0.240 & -28.7 & 30.2 & 14.4 & 0.5234 \\
208.75 & 67.90 & 0.05 & 0.157 & 4.7 & 6.6 & 3.4 & 0.4482 \\
208.90 & 75.99 & 0.14 & 0.195 & -12.1 & 16.8 & 10.6 & 0.9155 \\
209.06 & 80.28 & 0.21 & 0.207 & 5.8 & 8.5 & 16.9 & 0.3828 \\
209.11 & 43.68 & 0.31 & 0.245 & -13.5 & 17.4 & 13.5 & 0.1352 \\
209.40 & 113.19 & 0.08 & 0.151 & -5.6 & 5.6 & 9.1 & 0.4520 \\
209.41 & 74.34 & 0.11 & 0.233 & -8.6 & 9.9 & 8.2 & 0.6402 \\
209.45 & 48.10 & 0.35 & 0.343 & -15.7 & 24.3 & 16.8 & 0.8441 \\
209.49 & 83.57 & 0.16 & 0.263 & 6.1 & 8.3 & 13.4 & 0.4302 \\
209.72 & 58.71 & 0.10 & 0.245 & 4.7 & 12.7 & 5.9 & 0.9352 \\
209.80 & 43.51 & 0.40 & 0.276 & 1.9 & 24.9 & 17.4 & 0.6658 \\
209.85 & 84.25 & 0.09 & 0.209 & 14.3 & 14.3 & 7.6 & 0.8757 \\
210.06 & 87.87 & 0.08 & 0.175 & 1.2 & 5.0 & 7.0 & 0.4183 \\
210.23 & 82.40 & 0.08 & 0.187 & 1.4 & 5.5 & 6.6 & 0.6975 \\
210.28 & 53.02 & 0.08 & 0.205 & -7.1 & 7.1 & 4.2 & 0.0737 \\
210.33 & 31.34 & 0.14 & 0.134 & -1.5 & 7.4 & 4.4 & 0.6757 \\
210.42 & 77.14 & 0.06 & 0.134 & 4.1 & 6.7 & 4.6 & 0.0519 \\
210.49 & 85.55 & 0.06 & 0.172 & -2.5 & 4.4 & 5.1 & 0.5786 \\
210.57 & 56.21 & 0.09 & 0.147 & 2.8 & 3.7 & 5.1 & 0.2935 \\
210.58 & 35.45 & 0.21 & 0.206 & -1.1 & 13.5 & 7.4 & 0.8202 \\
210.68 & 64.91 & 0.18 & 0.240 & 4.2 & 10.7 & 11.7 & 0.4598 \\
210.69 & 83.23 & 0.14 & 0.192 & 12.6 & 14.5 & 11.7 & 0.3845 \\
210.73 & 21.43 & 0.20 & 0.223 & -4.2 & 8.1 & 4.3 & 0.9113 \\
210.85 & 81.06 & 0.23 & 0.282 & -6.1 & 9.4 & 18.6 & 0.9271 \\
211.04 & 58.17 & 0.06 & 0.136 & -14.6 & 14.8 & 3.5 & 0.7172 \\
211.14 & 94.67 & 0.09 & 0.144 & 3.3 & 3.5 & 8.5 & 0.5825 \\
211.20 & 39.55 & 0.09 & 0.173 & 0.4 & 3.2 & 3.6 & 0.0340 \\
211.29 & 86.71 & 0.09 & 0.198 & -8.0 & 8.0 & 7.8 & 0.7112 \\
211.50 & 82.50 & 0.14 & 0.200 & -0.9 & 9.7 & 11.5 & 0.4795 \\
211.60 & 38.73 & 0.27 & 0.281 & 8.7 & 13.4 & 10.5 & 0.7211 \\
211.63 & 75.81 & 0.10 & 0.188 & 8.6 & 10.1 & 7.6 & 0.3072 \\
211.76 & 74.33 & 0.07 & 0.148 & 5.1 & 8.5 & 5.2 & 0.1725 \\
211.91 & 78.05 & 0.13 & 0.255 & -3.7 & 9.2 & 10.1 & 0.5270 \\
211.99 & 92.71 & 0.04 & 0.194 & 4.5 & 4.5 & 3.7 & 0.7903 \\
212.06 & 71.46 & 0.12 & 0.294 & 15.2 & 17.1 & 8.6 & 0.6180 \\
212.06 & 92.56 & 0.13 & 0.200 & 13.5 & 13.5 & 12.0 & 0.6557 \\
212.06 & 40.96 & 0.12 & 0.206 & -1.7 & 9.2 & 4.9 & 0.6180 \\
212.20 & 29.21 & 0.13 & 0.210 & 5.9 & 12.2 & 3.8 & 0.7091 \\
212.27 & 61.24 & 0.11 & 0.218 & -10.5 & 10.8 & 6.7 & 0.1606 \\
212.31 & 125.72 & 0.03 & 0.124 & -2.0 & 2.6 & 3.8 & 0.1606 \\
212.47 & 53.67 & 0.07 & 0.132 & -5.4 & 7.7 & 3.8 & 0.5150 \\
212.48 & 74.77 & 0.05 & 0.134 & -2.6 & 7.1 & 3.7 & 0.5150 \\
212.56 & 30.20 & 0.11 & 0.176 & 8.8 & 13.1 & 3.3 & 0.3427 \\
212.73 & 70.71 & 0.06 & 0.113 & 0.7 & 3.1 & 4.2 & 0.4338 \\
212.83 & 107.27 & 0.09 & 0.156 & 6.4 & 8.8 & 9.7 & 0.2991 \\
212.95 & 112.81 & 0.03 & 0.105 & 2.5 & 2.9 & 3.4 & 0.9763 \\
213.10 & 85.00 & 0.05 & 0.153 & 1.4 & 3.3 & 4.3 & 0.1803 \\
213.17 & 113.55 & 0.03 & 0.120 & -0.6 & 2.1 & 3.4 & 0.6318 \\
213.22 & 31.60 & 0.17 & 0.194 & 2.5 & 7.8 & 5.4 & 0.5347 \\
213.27 & 75.28 & 0.06 & 0.154 & -7.5 & 7.7 & 4.5 & 0.4971 \\
213.28 & 93.30 & 0.08 & 0.145 & -7.2 & 8.7 & 7.5 & 0.6476 \\
213.34 & 49.40 & 0.10 & 0.197 & 0.2 & 10.6 & 4.9 & 0.1743 \\
213.48 & 74.65 & 0.09 & 0.167 & -15.5 & 15.5 & 6.7 & 0.1901 \\
213.57 & 30.07 & 0.39 & 0.226 & -1.0 & 9.1 & 11.7 & 0.9960 \\
213.79 & 78.39 & 0.09 & 0.177 & 7.5 & 10.7 & 7.1 & 0.5228 \\
213.88 & 48.02 & 0.16 & 0.203 & -3.6 & 17.5 & 7.7 & 0.4633 \\
214.05 & 69.27 & 0.06 & 0.153 & 0.7 & 2.4 & 4.2 & 0.3663 \\
214.09 & 93.47 & 0.04 & 0.137 & -1.3 & 3.8 & 3.7 & 0.5920 \\
214.10 & 26.08 & 0.13 & 0.158 & 4.7 & 5.8 & 3.4 & 0.0059 \\
214.24 & 68.63 & 0.10 & 0.170 & -4.9 & 6.7 & 6.9 & 0.9465 \\
214.40 & 31.81 & 0.13 & 0.234 & 2.8 & 6.2 & 4.1 & 0.2633 \\
214.44 & 114.29 & 0.03 & 0.146 & -0.9 & 1.8 & 3.4 & 0.1880 \\
215.10 & 73.51 & 0.29 & 0.215 & -1.2 & 14.9 & 21.3 & 0.1325 \\
215.10 & 34.93 & 0.14 & 0.244 & -16.3 & 16.3 & 4.9 & 0.5681 \\
215.13 & 90.47 & 0.04 & 0.161 & -0.5 & 1.0 & 3.6 & 0.4176 \\
215.27 & 50.53 & 0.06 & 0.225 & -8.0 & 8.6 & 3.0 & 0.5463 \\
215.28 & 83.08 & 0.05 & 0.182 & 0.3 & 4.8 & 4.2 & 0.5840 \\
215.32 & 99.75 & 0.05 & 0.185 & 4.1 & 7.1 & 5.0 & 0.8849 \\
215.38 & 108.22 & 0.03 & 0.128 & 2.8 & 5.5 & 3.2 & 0.2612 \\
215.57 & 28.69 & 0.17 & 0.229 & -16.1 & 19.0 & 4.9 & 0.2176 \\
215.84 & 29.17 & 0.11 & 0.188 & 12.6 & 13.5 & 3.2 & 0.0235 \\
215.87 & 76.25 & 0.38 & 0.233 & 29.0 & 30.3 & 29.0 & 0.2651 \\
216.03 & 55.02 & 0.06 & 0.142 & 10.3 & 12.3 & 3.3 & 0.2651 \\
216.06 & 20.01 & 0.15 & 0.182 & -18.3 & 18.3 & 3.0 & 0.8294 \\
216.23 & 36.97 & 0.12 & 0.190 & 2.4 & 6.3 & 4.4 & 0.9957 \\
216.29 & 53.74 & 0.11 & 0.176 & -6.6 & 13.3 & 5.9 & 0.4096 \\
216.32 & 82.50 & 0.23 & 0.199 & 2.4 & 15.0 & 19.0 & 0.0868 \\
216.38 & 65.20 & 0.05 & 0.199 & -8.2 & 8.6 & 3.3 & 0.8611 \\
216.49 & 34.45 & 0.32 & 0.243 & -9.5 & 16.8 & 11.0 & 0.7046 \\
216.60 & 67.67 & 0.07 & 0.134 & -1.2 & 6.1 & 4.7 & 0.5917 \\
216.69 & 77.51 & 0.11 & 0.181 & -2.8 & 8.0 & 8.5 & 0.4194 \\
216.77 & 50.07 & 0.14 & 0.203 & 7.8 & 13.3 & 7.0 & 0.1343 \\
217.01 & 74.49 & 0.07 & 0.156 & 1.5 & 5.2 & 5.2 & 0.6768 \\
217.22 & 34.25 & 0.09 & 0.236 & 1.7 & 10.0 & 3.1 & 0.3322 \\
217.27 & 83.69 & 0.06 & 0.127 & 1.5 & 2.1 & 5.0 & 0.5956 \\
217.33 & 33.22 & 0.20 & 0.154 & 19.2 & 28.9 & 6.6 & 0.5738 \\
217.34 & 104.14 & 0.05 & 0.126 & -1.6 & 2.5 & 5.2 & 0.0847 \\
217.49 & 94.47 & 0.06 & 0.203 & -13.3 & 13.3 & 5.7 & 0.2510 \\
217.49 & 72.58 & 0.10 & 0.157 & 16.5 & 16.5 & 7.3 & 0.4015 \\
217.81 & 53.69 & 0.09 & 0.204 & 6.9 & 7.9 & 4.8 & 0.7718 \\
217.86 & 24.36 & 0.18 & 0.225 & -5.1 & 9.0 & 4.4 & 0.4866 \\
217.93 & 84.03 & 0.04 & 0.158 & -3.9 & 5.4 & 3.4 & 0.4866 \\
217.96 & 55.08 & 0.08 & 0.186 & -1.7 & 4.2 & 4.4 & 0.8628 \\
218.16 & 80.59 & 0.06 & 0.138 & -4.8 & 4.8 & 4.8 & 0.2925 \\
218.23 & 87.46 & 0.09 & 0.183 & 0.3 & 5.6 & 7.9 & 0.9321 \\
218.33 & 35.80 & 0.12 & 0.158 & -4.1 & 9.7 & 4.3 & 0.7974 \\
218.54 & 64.52 & 0.05 & 0.154 & -0.8 & 4.2 & 3.2 & 0.1142 \\
218.55 & 83.58 & 0.08 & 0.152 & -2.9 & 2.9 & 6.7 & 0.3024 \\
218.57 & 28.61 & 0.11 & 0.158 & 6.1 & 8.0 & 3.1 & 0.5657 \\
218.61 & 64.64 & 0.10 & 0.155 & -11.2 & 11.6 & 6.5 & 0.8291 \\
218.86 & 69.32 & 0.11 & 0.171 & 3.6 & 5.8 & 7.6 & 0.7479 \\
219.05 & 46.49 & 0.15 & 0.159 & 1.3 & 6.2 & 7.0 & 0.3280 \\
219.09 & 73.93 & 0.12 & 0.159 & 0.9 & 6.3 & 8.9 & 0.5161 \\
219.33 & 56.49 & 0.13 & 0.323 & -13.2 & 13.2 & 7.3 & 0.3597 \\
219.49 & 48.45 & 0.08 & 0.177 & 9.8 & 10.5 & 3.9 & 0.3755 \\
219.50 & 77.14 & 0.26 & 0.179 & -29.6 & 33.1 & 20.1 & 0.1280 \\
219.64 & 39.34 & 0.10 & 0.159 & -1.9 & 6.9 & 3.9 & 0.5794 \\
219.65 & 89.64 & 0.05 & 0.141 & -0.0 & 1.4 & 4.5 & 0.4666 \\
219.71 & 82.80 & 0.04 & 0.172 & -0.9 & 2.2 & 3.3 & 0.8804 \\
219.77 & 76.43 & 0.09 & 0.175 & -8.7 & 8.7 & 6.9 & 0.5200 \\
219.78 & 54.37 & 0.09 & 0.160 & 2.3 & 4.2 & 4.9 & 0.5953 \\
219.95 & 77.55 & 0.04 & 0.135 & -7.1 & 7.1 & 3.1 & 0.6863 \\
219.96 & 63.77 & 0.08 & 0.162 & -6.2 & 7.1 & 5.1 & 0.9121 \\
220.01 & 79.61 & 0.07 & 0.178 & 3.2 & 5.5 & 5.6 & 0.2507 \\
220.09 & 35.64 & 0.10 & 0.151 & 11.0 & 16.2 & 3.6 & 0.9655 \\
220.10 & 86.24 & 0.05 & 0.138 & -0.6 & 1.9 & 4.3 & 0.8527 \\
220.27 & 33.88 & 0.13 & 0.184 & -16.2 & 20.0 & 4.4 & 0.4328 \\
220.43 & 25.88 & 0.13 & 0.207 & -7.1 & 12.2 & 3.4 & 0.6368 \\
220.53 & 64.27 & 0.05 & 0.163 & -1.7 & 3.5 & 3.2 & 0.0882 \\
220.59 & 126.11 & 0.06 & 0.116 & 2.4 & 3.6 & 7.6 & 0.5773 \\
220.62 & 32.23 & 0.20 & 0.156 & 5.7 & 14.3 & 6.4 & 0.3298 \\
220.73 & 113.46 & 0.03 & 0.108 & -2.7 & 5.3 & 3.4 & 0.5179 \\
220.78 & 75.78 & 0.17 & 0.166 & 11.7 & 12.7 & 12.9 & 0.4209 \\
\hline
\end{longtable}

\clearpage
\newpage

\begin{longtable}[c]{p{1.2cm}rccrrrrl} 
\caption{Detailed information of every frequency group found, made from frequencies with very high power (P) in the power spectrum, $P \ge 0.15$ in this case. The columns are the same as the previous Table \ref{tab:table3}. The stricter filter means that there are less groups in general, but the ones that remain represent the most relevant frequencies. }
\label{tab:table4}\\
\hline
\begin{tabular}[l]{@{}l@{}}Starting \\ date (JD)\\  \end{tabular} & \begin{tabular}[l]{@{}l@{}} $\overline{f}$ (d$^{-1}$)\end{tabular} & \begin{tabular}[l]{@{}l@{}}Duration\\  (JD)\end{tabular} & \begin{tabular}[l]{@{}l@{}}$\overline{P}$\end{tabular} & \begin{tabular}[l]{@{}l@{}}$f_{\Delta}$ (d$^{-1}$)\end{tabular} & \begin{tabular}[l]{@{}l@{}}$f_{A}$ (d$^{-1}$) \end{tabular} & \begin{tabular}[l]{@{}l@{}}$N_{c}$\end{tabular} & Phase \\ 
\hline
\endfirsthead
\multicolumn{8}{c}%
{{\bfseries Table \thetable\ continued from previous page}} \\
\hline
\begin{tabular}[l]{@{}l@{}}Starting \\ date   (JD)\end{tabular} & \begin{tabular}[l]{@{}l@{}}$\overline{f}$ (d$^{-1}$)\end{tabular} & \begin{tabular}[l]{@{}l@{}}Duration\\  (JD)\end{tabular} & \begin{tabular}[l]{@{}l@{}}$\overline{P}$\end{tabular} & \begin{tabular}[l]{@{}l@{}}$f_{\Delta}$ (d$^{-1}$)\end{tabular} & \begin{tabular}[l]{@{}l@{}}$f_{A}$ (d$^{-1}$) \end{tabular} & \begin{tabular}[l]{@{}l@{}}$N_{c}$\end{tabular} & Phase \\ 
\hline
\endhead
\hline
\endfoot
\endlastfoot
206.06 & 100.52 & 0.03 & 0.221 & 0.4 & 2.5 & 3.0 & 0.1317 \\
206.12 & 44.10 & 0.15 & 0.422 & -7.3 & 16.2 & 6.6 & 0.0346 \\
206.29 & 62.82 & 0.09 & 0.241 & 4.5 & 7.2 & 5.7 & 0.0881 \\
206.30 & 42.31 & 0.12 & 0.207 & -2.5 & 7.7 & 5.1 & 0.2762 \\
206.57 & 86.22 & 0.10 & 0.217 & 7.0 & 9.4 & 8.6 & 0.2326 \\
206.63 & 51.15 & 0.34 & 0.422 & 10.8 & 22.6 & 17.4 & 0.5870 \\
207.06 & 50.87 & 0.06 & 0.206 & 3.4 & 5.4 & 3.1 & 0.7692 \\
207.10 & 114.66 & 0.03 & 0.181 & -0.9 & 1.7 & 3.4 & 0.9573 \\
207.12 & 95.72 & 0.04 & 0.208 & -0.5 & 1.4 & 3.8 & 0.1454 \\
207.15 & 44.01 & 0.30 & 0.280 & 1.7 & 17.6 & 13.2 & 0.3493 \\
207.29 & 114.54 & 0.03 & 0.188 & 3.5 & 3.8 & 3.4 & 0.3870 \\
207.53 & 49.84 & 0.15 & 0.346 & -14.5 & 17.0 & 7.5 & 0.6444 \\
207.59 & 72.95 & 0.09 & 0.222 & 10.8 & 10.8 & 6.6 & 0.8701 \\
207.75 & 43.97 & 0.17 & 0.405 & 1.9 & 11.7 & 7.5 & 0.3750 \\
208.01 & 60.85 & 0.05 & 0.230 & 1.0 & 4.5 & 3.0 & 0.8799 \\
208.06 & 54.03 & 0.06 & 0.215 & 5.9 & 9.2 & 3.2 & 0.2938 \\
208.09 & 43.26 & 0.16 & 0.343 & 9.5 & 13.9 & 6.9 & 0.8958 \\
208.21 & 66.68 & 0.07 & 0.195 & -4.6 & 7.4 & 4.7 & 0.4601 \\
208.38 & 56.02 & 0.21 & 0.361 & -8.9 & 19.1 & 11.8 & 0.2660 \\
208.41 & 86.85 & 0.14 & 0.383 & -1.0 & 6.0 & 12.2 & 0.2284 \\
208.67 & 48.54 & 0.37 & 0.278 & -11.7 & 21.6 & 18.0 & 0.0501 \\
208.74 & 81.95 & 0.12 & 0.288 & -22.0 & 22.0 & 9.8 & 0.6363 \\
208.90 & 83.32 & 0.05 & 0.215 & -4.8 & 6.1 & 4.2 & 0.5769 \\
209.07 & 80.10 & 0.16 & 0.227 & 6.3 & 7.7 & 12.8 & 0.2699 \\
209.26 & 40.36 & 0.14 & 0.330 & 0.1 & 12.9 & 5.7 & 0.6243 \\
209.43 & 74.30 & 0.07 & 0.285 & -6.6 & 7.8 & 5.2 & 0.6402 \\
209.43 & 112.94 & 0.03 & 0.187 & 0.3 & 0.6 & 3.4 & 0.4897 \\
209.45 & 48.10 & 0.35 & 0.343 & -15.7 & 24.3 & 16.8 & 0.8441 \\
209.54 & 82.99 & 0.10 & 0.329 & -2.4 & 7.0 & 8.3 & 0.5807 \\
209.73 & 58.13 & 0.08 & 0.272 & 1.6 & 12.5 & 4.7 & 0.9352 \\
209.85 & 84.25 & 0.09 & 0.209 & 14.3 & 14.3 & 7.6 & 0.8757 \\
209.98 & 47.47 & 0.21 & 0.312 & 7.7 & 11.1 & 10.0 & 0.3054 \\
210.01 & 68.66 & 0.06 & 0.187 & -2.7 & 4.6 & 4.1 & 0.9668 \\
210.24 & 82.76 & 0.06 & 0.211 & -0.5 & 5.5 & 5.0 & 0.6975 \\
210.29 & 52.48 & 0.07 & 0.218 & -2.5 & 5.1 & 3.7 & 0.1113 \\
210.59 & 32.49 & 0.10 & 0.226 & -1.5 & 4.7 & 3.2 & 0.4816 \\
210.69 & 64.79 & 0.15 & 0.261 & 0.6 & 10.7 & 9.7 & 0.4222 \\
210.72 & 83.01 & 0.09 & 0.218 & 5.8 & 9.4 & 7.5 & 0.4222 \\
210.75 & 21.39 & 0.16 & 0.246 & -0.6 & 8.1 & 3.4 & 0.9113 \\
210.87 & 80.91 & 0.20 & 0.304 & -8.7 & 9.4 & 16.2 & 0.9647 \\
211.29 & 87.03 & 0.08 & 0.206 & -6.0 & 6.3 & 7.0 & 0.6736 \\
211.51 & 82.14 & 0.11 & 0.216 & 9.7 & 9.7 & 9.0 & 0.4419 \\
211.63 & 37.70 & 0.09 & 0.271 & -10.2 & 13.4 & 3.4 & 0.2696 \\
211.63 & 73.26 & 0.06 & 0.217 & 6.8 & 7.6 & 4.4 & 0.1567 \\
211.73 & 43.54 & 0.12 & 0.349 & -11.5 & 17.7 & 5.2 & 0.1349 \\
211.92 & 78.18 & 0.12 & 0.267 & -6.7 & 9.2 & 9.4 & 0.5646 \\
211.99 & 92.71 & 0.04 & 0.194 & 4.5 & 4.5 & 3.7 & 0.7903 \\
212.06 & 70.54 & 0.11 & 0.306 & 13.3 & 15.1 & 7.8 & 0.5804 \\
212.06 & 91.00 & 0.08 & 0.238 & 12.3 & 13.5 & 7.3 & 0.4675 \\
212.06 & 40.71 & 0.10 & 0.219 & -0.1 & 8.6 & 4.1 & 0.5428 \\
212.27 & 61.50 & 0.10 & 0.226 & -10.0 & 10.8 & 6.1 & 0.1230 \\
212.84 & 107.45 & 0.06 & 0.169 & 6.0 & 7.8 & 6.4 & 0.2615 \\
213.24 & 31.47 & 0.10 & 0.225 & 6.8 & 7.8 & 3.1 & 0.4218 \\
213.37 & 50.21 & 0.06 & 0.242 & -11.9 & 15.0 & 3.0 & 0.2496 \\
213.49 & 75.42 & 0.07 & 0.179 & -9.6 & 9.8 & 5.3 & 0.1901 \\
213.80 & 30.07 & 0.12 & 0.334 & 1.8 & 8.1 & 3.6 & 0.7109 \\
213.81 & 79.57 & 0.06 & 0.197 & 5.4 & 7.9 & 4.8 & 0.5604 \\
213.88 & 51.75 & 0.09 & 0.236 & 5.5 & 9.4 & 4.7 & 0.2000 \\
214.24 & 70.52 & 0.05 & 0.189 & -0.0 & 1.8 & 3.5 & 0.7584 \\
215.10 & 74.80 & 0.08 & 0.194 & -3.5 & 7.4 & 6.0 & 0.3424 \\
215.10 & 34.93 & 0.14 & 0.244 & -16.3 & 16.3 & 4.9 & 0.5681 \\
215.22 & 72.83 & 0.12 & 0.274 & -10.7 & 11.9 & 8.7 & 0.3958 \\
215.28 & 83.01 & 0.04 & 0.190 & -0.2 & 4.8 & 3.3 & 0.5463 \\
215.58 & 28.68 & 0.12 & 0.268 & -12.7 & 14.5 & 3.4 & 0.1047 \\
215.89 & 66.03 & 0.14 & 0.281 & 4.9 & 11.5 & 9.2 & 0.5126 \\
216.12 & 87.04 & 0.13 & 0.235 & 7.6 & 9.1 & 11.3 & 0.2056 \\
216.34 & 81.14 & 0.18 & 0.216 & 14.1 & 15.1 & 14.6 & 0.0492 \\
216.52 & 35.01 & 0.24 & 0.281 & -10.8 & 13.4 & 8.4 & 0.6293 \\
216.69 & 78.66 & 0.07 & 0.211 & 1.7 & 5.2 & 5.5 & 0.2689 \\
216.81 & 48.90 & 0.08 & 0.250 & 10.0 & 10.4 & 3.9 & 0.2095 \\
217.50 & 93.52 & 0.05 & 0.216 & -11.4 & 11.4 & 4.7 & 0.2886 \\
217.82 & 52.60 & 0.06 & 0.237 & 4.3 & 4.7 & 3.2 & 0.7341 \\
217.90 & 24.71 & 0.14 & 0.249 & -5.2 & 9.0 & 3.5 & 0.6371 \\
217.98 & 54.82 & 0.06 & 0.203 & 0.0 & 4.2 & 3.3 & 0.9381 \\
218.24 & 88.47 & 0.05 & 0.221 & 4.6 & 4.6 & 4.4 & 0.8569 \\
218.88 & 69.61 & 0.07 & 0.198 & 4.6 & 4.6 & 4.9 & 0.7479 \\
219.35 & 55.83 & 0.11 & 0.354 & -9.8 & 12.5 & 6.1 & 0.4349 \\
219.56 & 78.33 & 0.14 & 0.203 & -17.6 & 18.1 & 11.0 & 0.1280 \\
219.77 & 77.96 & 0.06 & 0.194 & -1.3 & 2.4 & 4.7 & 0.4071 \\
220.03 & 79.78 & 0.04 & 0.201 & 5.4 & 5.4 & 3.2 & 0.2883\\
\hline
\end{longtable}

\end{appendix}
\end{document}